## Domain Wall Sliding-induced Polarization Switching in Multilayer Graphene

Zhou Zhou,[1,*] Xiyao Peng,[1,*] Jianfeng Bi,[1] Fei Xue,[1] Jie Jiang,[2] Huizhen Wu,[1] Zhiwen Shi,[3] Haoliang Qian,[4,†] Toshikaze Kariyado,[5,‡] Sihan Zhao[1,§]

[1]School of Physics, Zhejiang University, Zhejiang Key Laboratory of Micro-Nano Quantum Chips and Quantum Control, and State Key Laboratory of Silicon and Advanced Semiconductor Materials, Zhejiang University, Hangzhou 310058, China.
[2]School of Materials Science and Engineering, Zhejiang University, Hangzhou 310058, China.
[3]Key Laboratory of Artificial Structures and Quantum Control (Ministry of Education), School of Physics and Astronomy and Tsung-Dao Lee Institute, Shanghai Jiao Tong University, Shanghai 200240, China.
[4]State Key Laboratory of Extreme Photonics and Instrumentation, ZJU-Hangzhou Global Scientific and Technological Innovation Center, College of Information Science and Electronic Engineering, Zhejiang University, Hangzhou 310058, China.
[5]Research Center for Materials Nanoarchitectonics (MANA), National Institute for Materials Science (NIMS), 1-1 Namiki Tsukuba 305-0044, Japan.

[*]These authors contributed equally to this work.
[†]Contact author: haoliangqian@zju.edu.cn
[‡]Contact author: KARIYADO.Toshikaze@nims.go.jp
[§]Contact author: sihanzhao88@zju.edu.cn

**ABSTRACT**. Electric polarization and metallicity are long believed not to coexist until the emergence of exceptionally rare material examples including the bulk polar metals and more recently two-dimensional (2D) van der Waals (vdW) materials such as 1T' $WTe_2$. The electric polarization for the latter represents a new and distinguishable paradigm in materials science and physics because its electric polarization states embedded in the conduction electron sea are able to couple with (and controlled by) the external electric field. However, the microscopic polarization switching process and mechanism in these 2D vdW metallic materials have not been experimentally observed and remain elusive. Here, we report the first direct experimental imaging of the microscopic mechanism behind electric-field-coupled polarization switching in a metallic system. Our gate-tunable nanoscale optical imaging identifies the robust coexistence of electric polarization and appreciable carrier densities in adjacent polar domains hosting opposite electric polarizations in a 2D elemental metallic material, tetralayer graphene. We directly visualize and verify that the sliding domain wall (DW) solitons confined at the middlemost interface are responsible for the polarization switching in tetralayer graphene upon the application of electric fields and mechanical forces. Our work provides the first direct visualization of domain wall sliding-induced polarization switching in 2D elemental carbon at room temperature, significantly expanding and advancing the research of "ferroelectric metal" initially dubbed by P. W. Anderson and coauthors in 1965.

*Introduction*—Electric polarization and metallicity are conventionally regarded as two contradictory properties that are rarely observed within a single material due to the incompatible electric dipoles and conduction electrons. In 1965, P. W. Anderson and coauthors initially proposed a theoretical material concept dubbed "ferroelectric metal" that challenged this conventional wisdom [1]. Recent advancements in two-dimensional (2D) van der Waals (vdW) materials have unlocked tremendous new opportunities to explore new platforms and physics permitting electric-field switchable electric polarization in metallic systems in the 2D limit beyond the bulk polar metals, the first material class hosting both electric polarization and metallicity initially discovered in 2013 [2]. A new type of polarization switching mechanism, referred to as 2D sliding ferroelectricity [3], has been recently studied predominantly in insulating and/or semiconducting 2D vdW materials [4-17], where the out-of-plane electric field can couple with the in-plane atomic motion unconventionally, leading to switchable polarizations by domain wall (DW) sliding. However, in 2D vdW metallic phases, to date, only atomically thin 1T' $WTe_2$ [18-22], $T_d$-$MoTe_2$ [23], and graphene (multilayer graphene)-hBN moiré superlattices [24-28] have shown global transport features of current hysteresis across macroscopic samples, tentatively interpreted as signatures of ferroelectricity. Due to the absence of a direct visualization on the polarization switching process, it poses a significant debate on their origin and switching mechanism in these emergent 2D vdW metallic materials.

Rich stacking orders in natural graphene polytypes offer a new opportunity to microscopically explore the polarization switching phenomena in 2D vdW metallic phases by studying the purely stacking-induced electric polarization and its switching [29,30]. Among all the polytypes, thinnest possible multilayer graphene hosting the required broken inversion and/or mirror symmetries to exhibit electric polarization is the tetralayer graphene with ABAC and ABCB type of polar stackings [29-33]. Their identification in tetralayer graphene was recently reported using Kelvin probe force microscopy (KPFM) measurement [34], further promising the microscopic study of polarization switching phenomena and mechanism in 2D metallic systems by using tetralayer graphene.

Bernal (ABAB) and rhombohedral (ABCA) stackings (ABCA) are centrosymmetric [Fig. 1(a)], with the latter attracting much research attention in the intertwined electron correlation and topology [35-38]. In contrast, the ABAC and ABCB stackings orders with purely stacking-induced broken (out-of-plane) inversion and mirror symmetries [Fig. 1(a)] carry finite and opposite electric polarizations, which is also supported by the calculated layer-projected band structure and charge density distribution using density functional theory (DFT) (Supplemental Material, Fig. S1 and Sec. 1 [39]). The vertical polarization can be understood by the imbalanced charge density distribution across graphene layers where the greatest charge imbalance is found between the second layer (L2) and the third layer (L3) (Supplemental Material, Figs. S1 and S2).

The tetralayer graphene is prepared on a $SiO_2$/Si substrate with 285 nm $SiO_2$ by mechanical exfoliation. We use gate-tunable scanning near-field optical microscopy (SNOM) [schematic shown in Fig. 1(b)] to identify different stacking orders in tetralayer graphene (Supplemental Material, Sec. 2 [39]), which has recently been utilized to indirectly probe the out-of-plane polarizations of twisted $WSe_2$ and hBN moiré superlattices by using a graphene sensor [15,57-59]. Figure 1(c) shows the atomic force microscopy (AFM) topography of a typical tetralayer graphene with a uniform height, while the SNOM image of the same region shown in Fig. 1(d) reveals three regions with distinct near-field optical contrasts. The three graphene regions in Fig. 1(d) are individual domains of ABAB (black dot) and ABCA (blue dot) and a polar ABCB-type domain (denoted as ABCB* if not specifically distinguished by the red dot). This stacking assignment is supported by complementary Raman spectroscopy measurements on 2D peaks [Fig. 1(e)] [55,56]. Representative SNOM images and statistical area percentages of ABCB-type domains are shown in Supplemental Material, Fig. S3.

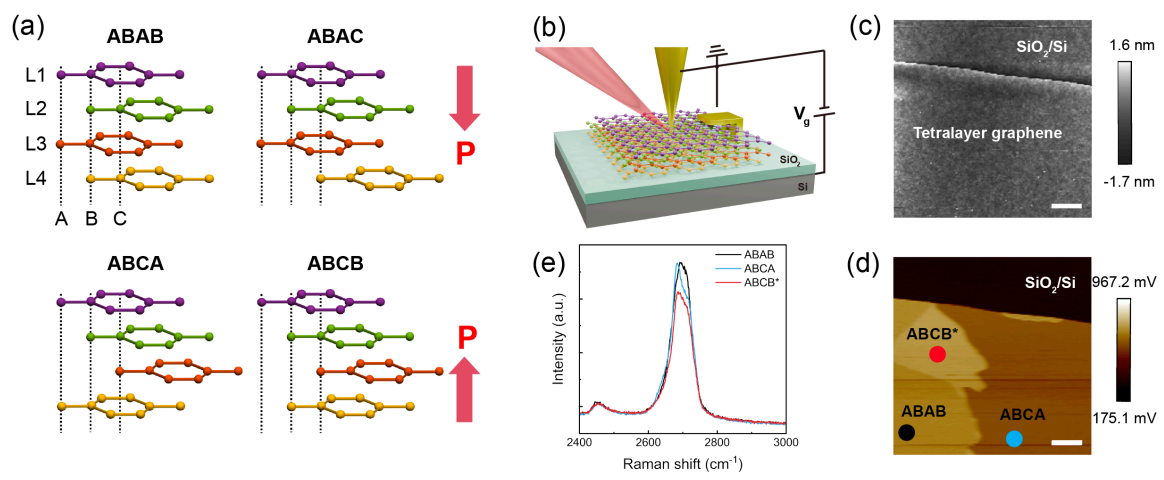

FIG. 1. Schematics and characterizations of tetralayer graphene with broken symmetries. (a) Stacking configurations of tetralayer graphene, including nonpolar ABAB and ABCA and polar ABAC and ABCB stackings. (b) Schematic of the gate-tunable SNOM measurement. (c), (d) Topology and SNOM images of a typical tetralayer graphene. The image in (d) shows three different stacking orders, corresponding to ABAB (black dot), ABCA (blue dot) and ABCB-type (denoted as ABCB*, red dot) domains. (e) Raman 2D peaks measured from the three different domains indicated by the colored dots in (d). The scale bars in (c) and (d) are 2 μm.

*Gate-tunable SNOM imaging of adjacent polar states with opposite polarizations*—To enable the direct visualization of microscopic polarization switching in tetralayer graphene, one challenge is to discern adjacent polar domains with opposite electric polarizations. We apply back-gate voltages ($V_g$) to induce electron and/or hole doping in tetralayer graphene [Fig. 1(b)] (Supplemental Material, Sec. 2 [39]). Figures 2(a)-2(h) present the representative gate-dependent SNOM images of a tetralayer graphene device (device 1), in which SNOM data measured from $V_g$ = 0 V to $V_g$ = 60 V and from 0 V to $V_g$ = -45 V are shown. At $V_g$ = 0 V [Fig. 2(a), close to charge neutrality], SNOM can distinguish Bernal (ABAB), rhombohedral (ABCA) and polar (ABCB*) stackings in the same way as demonstrated in Fig. 1.

We focus on the polar stacking region in the middle of each SNOM image [marked by I, II, and III in Fig. 2(c) and triangles in Fig. 2(b)]. As $V_g$ is increased from 0 V, a DW across the whole region becomes clearly visible, dividing the original region with nearly identical optical contrast into two parts, region I and region II. With positive $V_g$ (electron doping), the near-field response of region II becomes progressively darker than that of region I [Figs. 2(a)-2(d)]. However, it shows the exact opposite trend under negative $V_g$ [Figs. 2(e)-2(h)], in which the near-field response of region II becomes progressively brighter than that of region I (details in Supplemental Material, Sec. 3 [39]). Consistent behaviors with those in Figs. 2(a)-2(h) (first cycle scan) are observed in a consecutive gate-dependent SNOM scan (second cycle scan), with extension to a larger range of $V_g$. Full SNOM datasets for two scans are shown in Supplemental Material, Figs. S4 and S5.

Figure 2(i) quantifies the near-field optical contrasts between regions I and II $(S_I - S_{II})/S_{II}$ for the first cycle scan and between regions I and III $(S_I - S_{III})/S_{III}$ for the second cycle scan (region II diminished to very tiny area after the first cycle scan), where $S$ represents the measured near-field signal averaged over selected areas in each region (extraction details in Supplemental Material, Sec. 2 [39]). Supplemental Material, Figure S6 presents the full datasets for the extracted optical contrasts. We also confirm that the extracted gate-dependent optical contrasts show consistent trends, not depending on the areas selected (Supplemental Material, Fig. S7). We therefore achieve the optical detection of two polar stacking orders with opposite vertical electric polarizations. This is further verified by correlating the SNOM with KPFM results shown in Supplemental Material, Fig. S8, Sec. 4 [39], alongside the Raman spectra supporting the stacking assignments marked in Fig. 2(b). Our data in Fig. 2 also clearly indicates the coexistence of electric polarization and large number of free carriers at least at $\sim 5 \times 10^{12}\ cm^{-2}$ [Fig. 2(i)] in tetralayer graphene with broken symmetries.

Our second device (device 2) exhibits identical domain patterns as observed in device 1 (Fig. 2). The full gate-dependent SNOM dataset and the KPFM data, together with the Raman spectra of device 2, are shown in Supplemental Material, Figs. S9 and S10. We also systematically measure two extra devices (devices 3 and 4) via gate-dependent SNOM, KPFM, and Raman spectroscopy, obtaining consistent results (Supplemental Material, Figs. S11 to S16). The extracted optical contrasts via gate-dependent SNOM for devices 2 to 4 are displayed in Supplemental Material, Fig. S17. The gate-dependent near-field response of the polar stackings [Figs. 2(i) & S17] can be qualitatively understood by the carriers filling according to the calculated layer-projected charge distribution as detailed in Supplemental Material, Sec. 5 and Fig. S18.

*Polarization switching in gated tetralayer graphene*—Figures 3(a)-3(e) display representative SNOM snapshots for device 1, in which Figs. 3(a)-3(c) and Figs. 3(d) and 3(e) are obtained in the first cycle and second cycle scans, respectively (see also Supplemental Material, Figs. S4 and S5). We observe two distinct types of DWs confined at different interfaces in tetralayer graphene, which are indicated by the blue and green dashed lines in Figs. 3(a)-3(e),

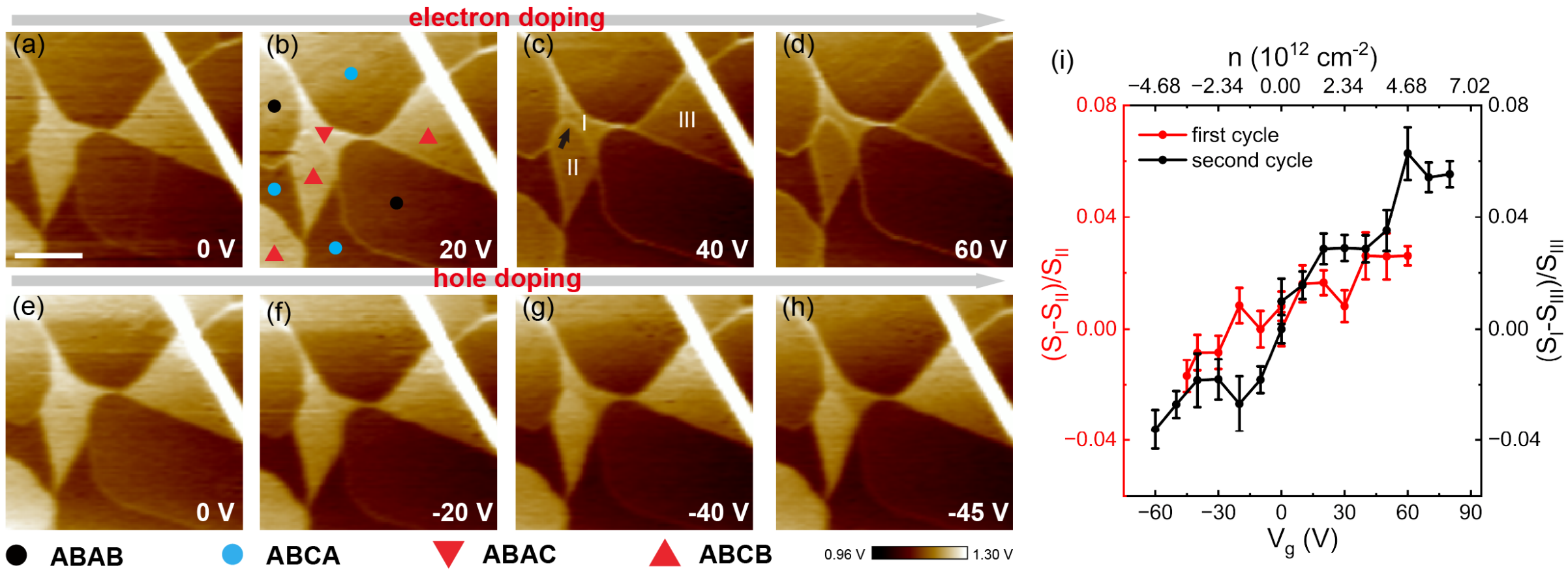

FIG. 2. Gate-tunable SNOM imaging of adjacent polar states with opposite polarizations for device 1. (a)-(h) SNOM images at representative back-gate voltages ($V_g$) for 0 V – 60 V [(a)-(d)] with electron doping and 0 V – -45 V [(e)-(h)] with hole doping. Domains with distinct stacking orders are denoted by different symbols in (b). A DW separates two adjacent polar stackings ABAC and ABCB, denoted as regions I and II in (c); regions III and II have the same polar stacking. The polarization directions are experimentally determined by an independent KPFM measurement [Supplemental Materials, Fig. S8(c)]. All the data in (a)-(h) are presented using the same color bar. The scale bar in (a) is 1 μm. (i) Extracted optical contrast between two polar stackings as a function of $V_g$ (bottom axis) and the carrier density ($n$, top axis). The data for $(S_I - S_{II})/S_{II}$ in the first cycle and for $(S_I - S_{III})/S_{III}$ in the second cycle measurements are displayed by the red and black dots, respectively.257

respectively, with arrows indicating the corresponding sliding directions. The schematic drawing of the domain patterns and DWs in Fig. 3(f) reproduces the domain patterns and pronounced DW sliding revealed in Figs. 3(a)-3(e). The blue solid line in Fig. 3(f) depicts the DW shared between polar CACB (ABAC) and CABA (ABCB), and between nonpolar ABAB and ABCA. The green solid line depicts another type of DW shared between polar CACB (ABAC) and nonpolar ABAB, and between polar CABA (ABCB) and nonpolar ABCA. One important observation is the direct visualization of DW sliding indicated by the blue dashed lines in Figs. 3(a)-3(e), which unambiguously demonstrates the DW sliding mechanism when one polar stacking in tetralayer graphene is switched to another with opposite out-of-plane polarizations.

The observed DW sliding of different types in Figs. 3(a)-3(e) rationalizes the formed domain patterns under investigation, which can be directly translated into interlayer sliding in tetralayer graphene as follows. The movement of the blue DW in Fig. 3(f) corresponds to simultaneous sliding of the bottom two layers with the top two layers fixed [changing from CABA (ABCB) to CACB (ABAC), and from ABCA to ABAB], as illustrated in Fig. 3(g); it can also correspond to sliding of the top two layers with the bottom two layers fixed, as illustrated in Supplemental Material, Fig. S19(a). These two scenarios are completely equivalent, with the DWs both confined between the middle two layers L2 and L3. Therefore, the sliding of DW solitons at middlemost interface governs polarization switching in tetralayer graphene.

On the other hand, the movement of the green DW in Fig. 3(f) corresponds to the sliding of the top three layers with respect to the bottom layer [Supplemental Material, Figs. S19(b) & S20(a)]. We note that sliding of the topmost layer with respect to the bottom three layers is energetically unfavorable and thus unlikely to happen (Supplemental Material, Fig. S21).

To gain more physical insights into the DW sliding dynamics, we theoretically investigate the pathways for interlayer sliding observed in our experiments [Figs. 3(a)-3(f)] using the nudged elastic band (NEB) method (see Supplemental Material, Sec. 1 [39]). According to the DFT-calculated switching pathways presented in Fig. 3(h) and Supplemental Material, Fig. S20(b), although the two polar stackings are metastable phases, the calculated energy barriers for the four interlayer sliding pathways [Fig. 3(g) & Fig. S20(a)] vary slightly, which agrees with our observations in Figs. 3(a)-3(f) (discussion details in Supplemental Material, Sec. 6 [39]).

*Polarization switching by a local electrical gating*—Under the global back-gate gating (Fig. 3), the DW sliding between two polar domains may be affected by the presence of nearby nonpolar domains as the same DW is shared by multiple domains. We therefore carry out a polarization switching experiment by applying a local electric field between the biased tip ($V_{tip}$) and two adjacent polar domains in the vicinity of the DW (see Supplemental Material, Sec. 2 [39]) [6]. Figure 4(a) shows the SNOM image of device 2, which exhibits identical domains and DW patterns as in device 1 shown in Fig. 2 and Fig. 3, but with larger size of polar domains. Figure 4(a) is obtained after the mechanical manipulation after Fig. 4(i) but before the application of a local electric field. After scanning the same area of the sample [red box in Fig. 4(b)] with $V_{tip} = 4\ V$ [Fig. 4(b)], $-4\ V$ [Fig. 4(c)] and $-6\ V$ [Fig. 4(d)], the DW between two polar states (indicated by the blue dashed lines) remains unchanged. After scanning the same region with $V_{tip} = 8\ V$ [Fig. 4(e)], the DW between two polar states moves, transforming ABCB to ABAC (forward switching process). The cyan dashed line in Fig. 4(e) marks the new DW position. This observation means that the electric field, pointing downward (positive $V_{tip}$) and localized between the biased metallic tip and the two polar domains, switches the upward dipole in ABCB to become the downward dipole in ABAC via a DW sliding. This is an energetically favorable process. We note that the other DW (indicated by the green dashed line) also moves [new DW position marked by the light green dashed line in Fig. 4(e)]. However, a reverse switching process between two polar domains is not observed when tip bias is set to $V_{tip} = -8\ V$ [Fig. 4(f)]. Instead, the area after scanning with $V_{tip} = -8\ V$ shows a dark contrast, without signs of a successful reverse switching. The absence of reverse switching process is also observed in device 1 under global gating (not shown). Such asymmetry may be understood by taking the extrinsic substrate effect into account as a recent theoretical work suggests [60]. Nevertheless, future experimental investigation and theoretical calculation are necessary to fully resolve this switching asymmetry.

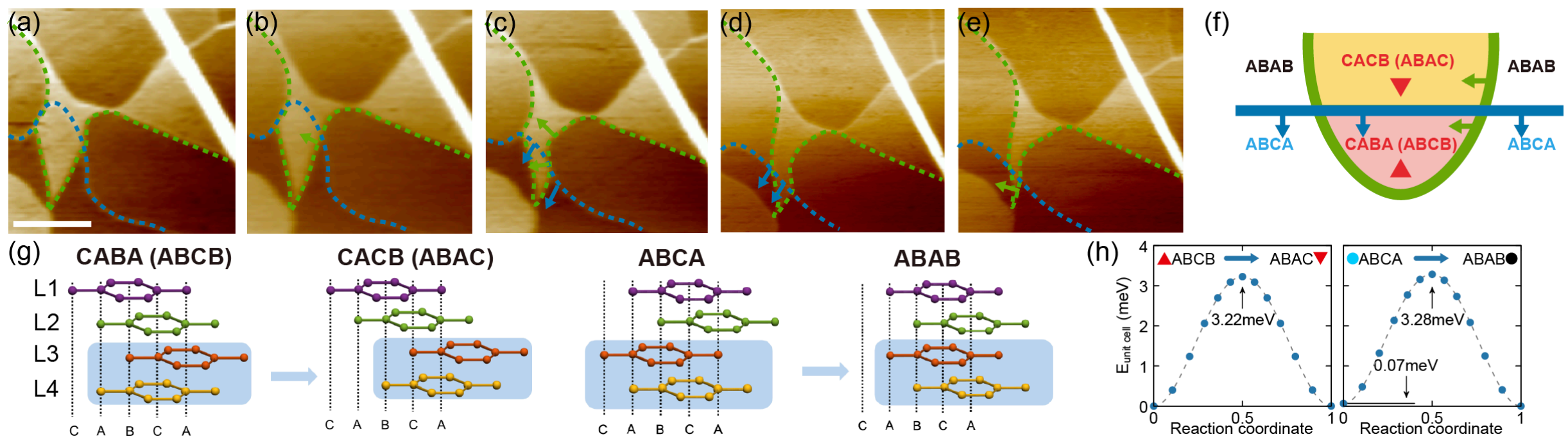

FIG. 3. Polarization switching in gated tetralayer graphene of device 1. (a)-(e) SNOM images capturing the sliding motions of two different types of DWs (indicated by the blue and green dashed lines, respectively) confined at different interfaces when gated. It enlarges the polar ABAC at the expense of polar ABCB; meanwhile it also enlarges the nonpolar ABAB at the expense of ABCA. The motion of the other type DW (indicated by the green dashed line) enlarges the nonpolar ABAB and ABCA stackings at the expense of the polar ABAC and ABCB stackings, respectively. The scale bar in (a) is 1 μm [same scale in (a)-(e)]. (f) A schematic reproducing the DW sliding and domain patterns observed in (a)-(e). (g) Schematic illustration for the blue DW motion in (f), corresponding to sliding of the bottom two layers together. (h) Calculated total energies for ABAB, ABCA, ABAC and ABCB stackings and energy barriers for the two switching pathways corresponding to the blue DW sliding observed in (a)-(e).207

*Polarization switching by mechanical tip motion*—We realize reversible switching between two adjacent polar ABAC and ABCB stackings by mechanically dragging the DW soliton with an AFM tip (see Supplemental Material, Sec. 2 [39]) [61], which supports and supplements the DW sliding-induced polarization switching mechanism demonstrated in Figs. 3 and 4. Figure 4(g) shows a SNOM image of device 2 prior to tip mechanical manipulation. The blue dashed line indicates the DW positions separating two polar stackings before tip manipulation. After the sample is scanned from left to right [Fig. 4(h)], the DW between two polar states is pulled to the right, transforming the ABAC stacking with downward polarization to ABCB stacking with upward polarization (reverse switching process). It is noteworthy that although the electric-field induced reverse switching process is not observed [Fig. 4(f)], such switching is indeed realized by the identical DW sliding mechanism. Figure 4(i) shows the resulting pattern after the tip is scanned in the backward direction following the movement shown in Fig. 4(h). The ABCB domain with upward polarization is partially transformed back into the ABAC domain with downward polarization (forward switching process), in which the new DW is depicted by the white dashed line. Therefore, electric polarizations in tetralayer graphene can be reversibly switched via mechanical manipulation of DW motions, albeit the switching is not fully recovered, possibly due to DW pinning at the ends and a slight mismatch between the two scanning areas in Figs. 4(h) and 4(i).

*Acknowledgements*—This work is supported mainly by the National Key R&D Program of China (2023YFA1407900 and 2022YFA1203400), the National Natural Science Foundation of China (12174335, 12374292, and 62175217), and the Zhejiang Provincial Natural Science Foundation of China (LR23A040002). TK was supported by JSPS

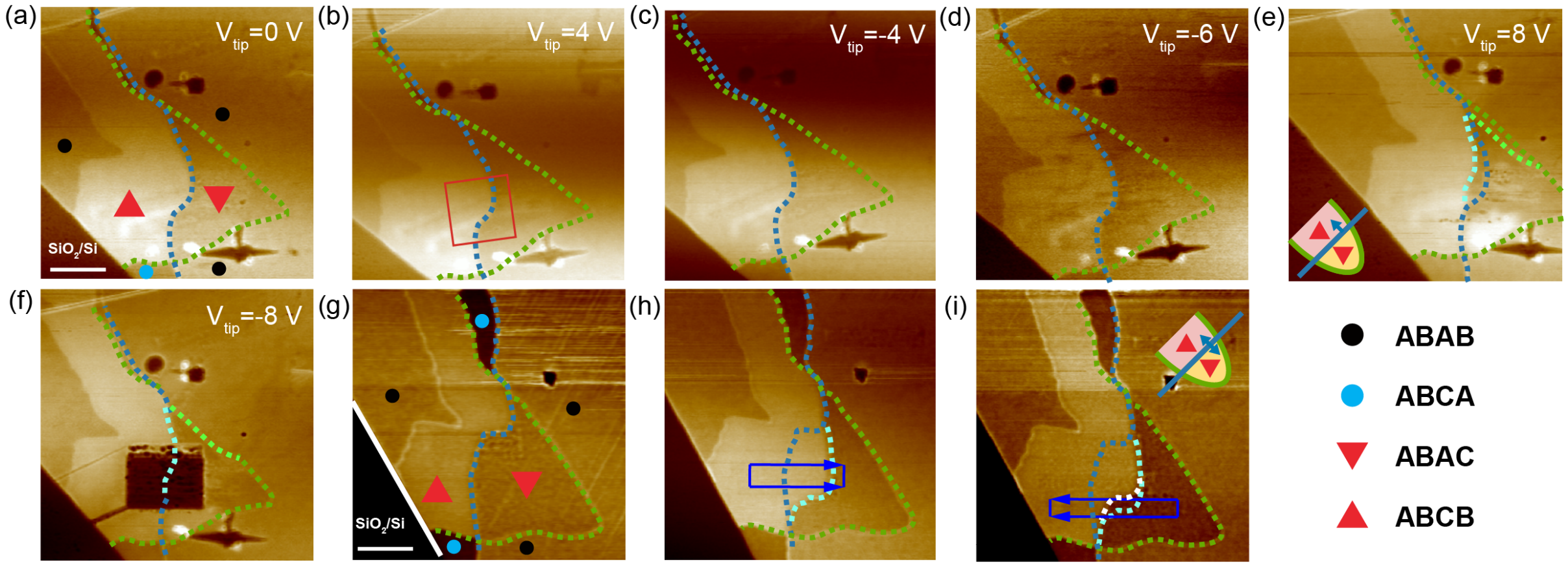

FIG. 4. Local electric field and mechanical switching between adjacent polar ABAC and ABCB stackings of device 2. (a)-(f) Sequential SNOM images after scanning a small area [red box in (b)] with $V_{tip} = 0, 4, -4, -6, 8,$ and $-8\,V$, respectively. Two DWs indicated by the blue and green dashed lines separate six domains whose stackings are denoted by symbols of different colors. The original DW between two polar states (blue dashed lines) moves at $V_{tip} = 8\,V$ (e), transforming ABCB to ABAC (new DW position indicated by the cyan dashed line). The other DW nearby also moves, with the new DW position indicated by the light green dashed line in (e). (g)-(i) Sequential SNOM images of device 2 before the tip mechanical manipulation (g), after a DW dragging to the right (h), and after a DW dragging to the left (i). The new DWs after the mechanical manipulation in (h) and (i) are indicated by the cyan and white dashed lines, respectively. The blue rectangles in (h) and (i) mark the scanning area of the AFM tip, with arrows indicating the tip movement directions. The insets of (e) and (i) show the schematics of the domain pattern and DW movement between polar states. The scale bar is 1 μm for all panels.

KAKENHI Grant Number JP24K06968. The calculations in this study were performed with the Numerical Materials Simulator at NIMS and using the facilities of the Supercomputer Center, the Institute for Solid State Physics, the University of Tokyo.

## End matter

*Polar metal*—Among the rare exceptions, bulk polar metals (including $LiOsO_3$, $Ca_3Ru_2O_7$, etc.) are the first class of materials reported in the 2010s where the two properties coexist, although long theorized by P. W. Anderson and coauthors since 1965 [1-2, 61]. This dual property in bulk polar metals is achieved through different mechanisms such as lattice distortion-induced phase transitions, carrier (dopant atom) doping to parent insulating ferroelectrics, etc [62-65]. While polar metals do show polarizations, experimental evidence for its polarization switching by external control knobs remains absent, except for a strain-induced ferroelastic switching study [62]. This fact manifests a fundamentally weak coupling effect between electric polarization and external electric field in bulk polar metals.

# Supplemental Material

# Domain Wall Sliding-induced Polarization Switching in Multilayer Graphene

Zhou Zhou,[1,*] Xiyao Peng,[1,*] Jianfeng Bi,[1] Fei Xue,[1] Jie Jiang,[2] Huizhen Wu,[1] Zhiwen Shi,[3] Haoliang Qian,[4,†] Toshikaze Kariyado,[5,‡] Sihan Zhao[1,§]

[1]School of Physics, Zhejiang University, Zhejiang Key Laboratory of Micro-Nano Quantum Chips and Quantum Control, and State Key Laboratory of Silicon and Advanced Semiconductor Materials, Zhejiang University, Hangzhou 310058, China.

[2]School of Materials Science and Engineering, Zhejiang University, Hangzhou 310058, China.

[3]Key Laboratory of Artificial Structures and Quantum Control (Ministry of Education), School of Physics and Astronomy and Tsung-Dao Lee Institute, Shanghai Jiao Tong University, Shanghai 200240, China.

[4]State Key Laboratory of Extreme Photonics and Instrumentation, ZJU-Hangzhou Global Scientific and Technological Innovation Center, College of Information Science and Electronic Engineering, Zhejiang University, Hangzhou 310058, China.

[5]Research Center for Materials Nanoarchitectonics (MANA), National Institute for Materials Science (NIMS), 1-1 Namiki Tsukuba 305-0044, Japan.

[*]These authors contributed equally to this work.

[†]Contact author: haoliangqian@zju.edu.cn

[‡]Contact author: KARIYADO.Toshikaze@nims.go.jp

[§]Contact author: sihanzhao88@zju.edu.cn

**Supplementary Figure 1.**

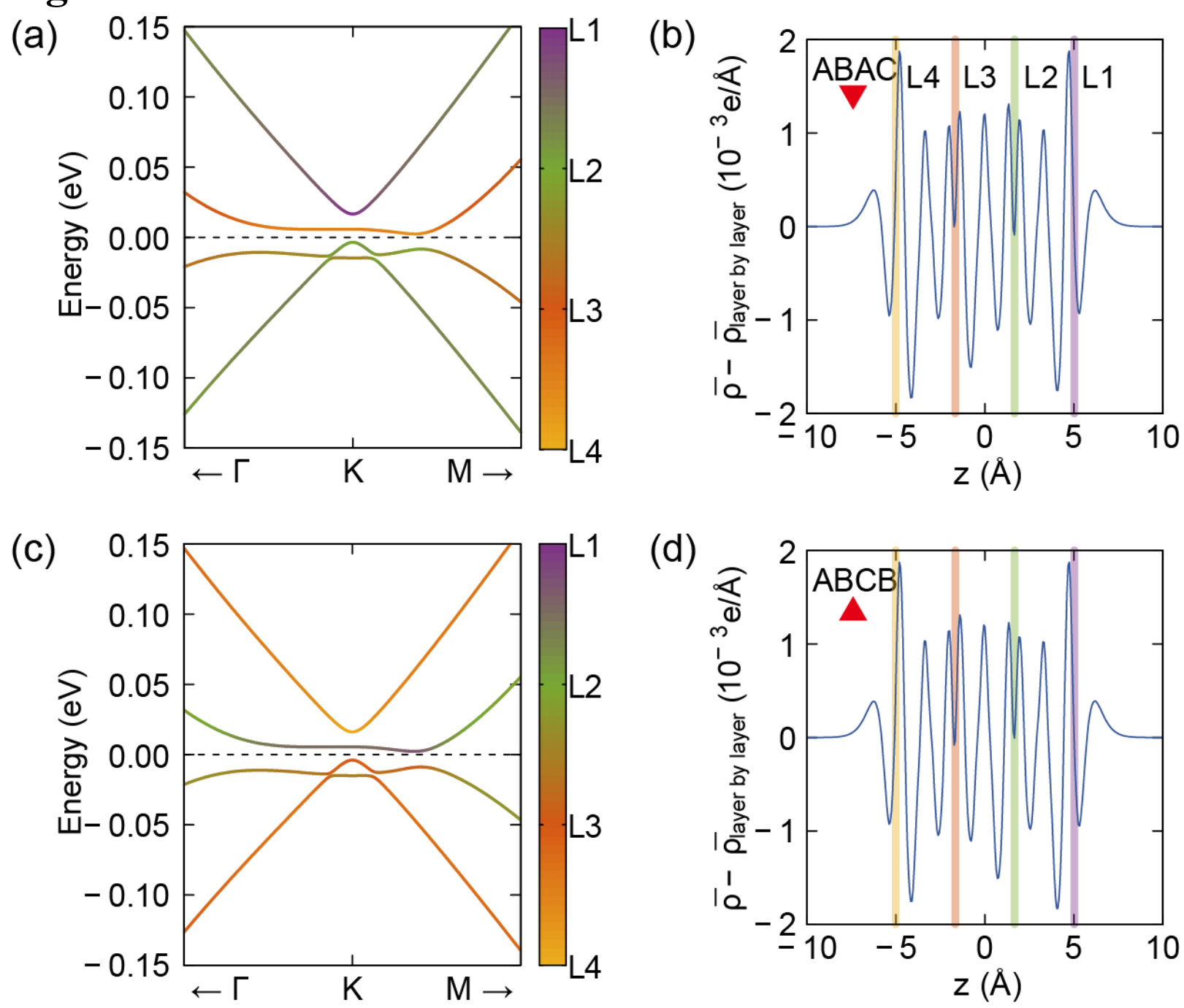

**FIG. S1. Calculated layer-projected band structure and charge density distribution for polar stackings.** (a),(b) Layer-projected band structure and charge density distribution for the polar ABAC stacking calculated by DFT. (c),(d) Layer-projected band structure and charge density distribution for the polar ABCB stacking calculated by DFT. For the band structures [(a) and (c)], each band is colored according to the weighted sum of the layer-resolved amplitude of the wavefunction. For the charge density [(b) and (d)], we subtract the superposition of the charge densities of the four independent graphene layers to better visualize the charge redistribution in the polar stacking order.

## Supplementary Note 1. Computational details

Density functional theory (DFT) calculations were performed via the Quantum Espresso package [1,2]. For structural relaxation, the rev-vdW-DF2 functional [3,4] was used to consider vdW interactions. After the structures were relaxed, the band structures and density of states (DOS) were calculated with the Perdew–Burke–Ernzerhof (PBE)-generalized gradient approximation (GGA) [5] functional (Fig. S1). We adapted pseudopotentials constructed with the projector augmented wave (PAW) method using parameters in PSlibrary [6,7]. The energy cutoffs of the wavefunctions and charge densities were 80 Ry and 640 Ry, respectively.

In self-consistent DFT calculations, k-points were sampled on $N \times N \times 1$ regular grids in which the in-plane grids were shifted by a half grid, and the convergence was checked up to $N = 264$. In the literature, there were found some variations in details of the band structure for ABCB/ABAC stackings. Mostly, the differences are explained by differences in theoretical or computational methods, e.g., a tight-binding model or DFT based calculations. However, already within the DFT, there is subtlety about convergence with respect to the momentum space k-mesh sampling. In practice, $N \times N$ in plane regular grids are used in the calculations. Note that the regular grids can be placed with half-grid shift to avoid the exact Γ-point or without such shift to include the exact Γ-point. The convergence is particularly slow without the half-grid shift, and the band structures look a little different for small k-sampling numbers (e.g., extra band crossing at the conduction band side).

In the DOS calculations, k-points were sampled in a small region near the K-point in the Brillouin zone (a circular region with an area of approximately 0.165% of the Brillouin zone area) to reduce the computational cost. For the charge distribution analysis, we calculated the partial Löwdin charge for each carbon atom (Fig. S2). Here, "partial" means that the integration over the Brillouin zone was limited to a small portion, as in the DOS calculations, since the important contributions were from the region near the K-point. Energy barriers between stacking patterns were estimated using the NEB method with 13 images along the reaction path.

**Supplementary Figure 2.**

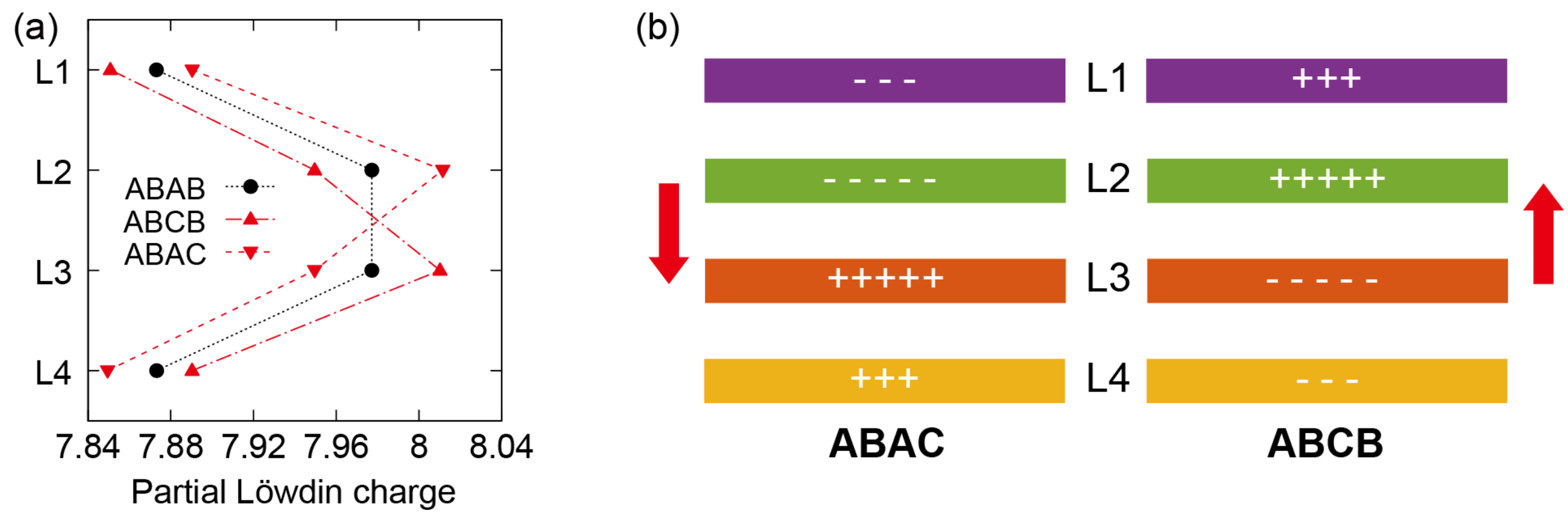

**FIG. S2. Partial Löwdin charge for each layer in the ABAC and ABCB stackings.** (a) Calculated partial Löwdin charge for each carbon atom. Considering the ABAB case as a reference (black dots and dotted line), there are fewer electrons in L1 and L2 and more electrons in L3 and L4 in the ABCB stacking, whereas there are more electrons in L1 and L2 and fewer electrons in L3 and L4 in the ABAC stacking. For both polar stackings, the charge imbalance, in comparison with that for ABAB, is greater in L2 and L3 than in L1 and L4. (b) Schematic drawing of the charge distributions for ABAC and ABCB according to (a). For ABAC stacking, more holes reside in L3 than in L2 [Fig. S1(b) and Fig. S2], carrying polarization direction pointing downward; whereas, on the other hand, more holes occupy L2 in the ABCB stacking, with polarization direction pointing upward [Fig. S1(d) and Fig. S2].

## Supplementary Note 2. Experimental Methods

Gate-tunable scanning near-field optical microscopy (SNOM)

The SNOM measurements were performed under ambient conditions. An infrared laser at ~ 10.6 μm was focused onto the apex of the gold-coated conductive AFM tip, which was operated in tapping mode (~ 200 kHz) during sample scanning. The signal scattered by the oscillating tip was subsequently detected by a high-speed HgCdTe detector and demodulated at the third harmonic of the tip tapping frequency by a lock-in amplifier to suppress the large back-reflected background. SNOM and topography images were simultaneously obtained. The gate-dependent SNOM measurements were carried out by following a closed gating loop, that is, we set the $V_g$ sweep from 0 V to the positive maximum, then from the positive maximum to the negative maximum, passing through 0 V, and then returned from the negative maximum back to 0 V.

Raman spectroscopy

Raman spectra were taken using a 532 nm laser focused by an objective (Nikon, with NA = 0.60). The backscattered signal was collected by the same objective and filtered before being dispersed by a 1200/mm grating spectrometer and detected by a liquid-$N_2$-cooled silicon detector (Princeton Instruments). The typical laser power and exposure time used were ~ 1.1 mW and ~ 60 s. The 2D peaks in graphitic multilayers, arising from a double-resonance Raman process, are sensitive probes of the local electronic band structure, and their spectral shape is widely used to distinguish different stacking orders in the literature [8-10]. The 2D peak of the ABAB stacking [black curve in Fig. 1(e) of the main text] has the most symmetric spectral shape, whereas that of the ABCA stacking [blue curve in Fig. 1(e) of the main text] has the most asymmetric spectral shape, with a strong peak feature on the lower wavenumber side. The spectral shape of the 2D peak of the polar stacking [red curve in Fig. 1(e) of the main text] lies between those of the blue and black curves. We note that the 2D peaks in Raman spectra cannot be used to distinguish the two polar stackings [11,12].

Kelvin probe force microscopy (KPFM)

KPFM measurements were conducted using a commercial scanning probe microscopy system (Dimension Icon, Bruker) in frequency-modulated (FM)-KPFM mode. Tips with a mechanical resonance frequency of ~ 75 kHz and a force constant of ~ 3 N/m were used. Areas with larger voltages in the KPFM mapping (brighter areas) correspond to stackings with lower work functions. For tetralayer graphene, the ABCA domains are the brightest, corresponding to the lowest work function. For the two nearby polar stackings indicated by downward and upward triangles, the former [downward triangle in Fig. S8(b) (region I in Fig.2 of the main text), Fig. S10(b), Fig. S12(b) and Fig. S15(b)] appears to be brighter than the latter [upward triangle in Fig. S8(b) (region II and region III in Fig.2 of the main text), Fig. S10(b), Fig. S12(b) and Fig. S15(b)]. This means that the two stackings have downward and upward polarizations. They are thus determined to be ABAC and ABCB stacking orders, respectively, according to our definition in Fig. 1(a) in the main text.

Near-field optical contrast between polar domains

To extract the optical contrast between, for example, regions I and II, i.e., $(S_I - S_{II})/S_{II}$, at each fixed back-gate voltage, we selected two small square areas in each domain (Figs. S6 and S7). We extracted $(S_I - S_{II})/S_{II}$ by using the near-field signals $S_I$ and $S_{II}$ averaged over two selected square areas in each domain. The error bar of $(S_I - S_{II})/S_{II}$ at each fixed back-gate voltage was computed by using the following formula [13]: $u = \left(\frac{S_I}{S_{II}}\right)^2 \cdot \left(\frac{u_1^2}{S_{II}^2} + \frac{u_2^2}{S_{II}^2}\right)$, where $u_I$ and $u_{II}$ are the standard deviations of $S_I$ and $S_{II}$, respectively. The carrier density $n$ at a fixed $V_g$ was estimated from the capacitance model $C = \frac{ne}{V_g} = \frac{\varepsilon_r \varepsilon_0}{d}$, where $d = 285$ nm and $\varepsilon_r = 4$ are the oxide thickness and dielectric constant, respectively, $e$ is the elementary charge, and $\varepsilon_0$ is the vacuum permittivity.

Electrical switching of polarization by a local electric field

To apply a local electric field to switch the polarization, the metallic AFM tip operated in a contact mode was biased with DC voltages while the tetralayer graphene sample was kept grounded. To rule out the possibility that the movement of the domain wall was caused by the tip operated in the contact mode itself, we scanned the same area under the same tip scanning condition without applying bias and we did not discern any DW movement by the contacted tip itself.

Mechanical manipulation of DWs with an AFM tip

The AFM tip was operated in normal tapping mode during forward scanning, but it was lifted down by ~ 0.3 μm (lift mode) during backward scanning with the feedback loop closed. Such scanning with a certain downward lift gave rise to a large lateral force that could slide the DW [14,15]. By rotating the scan area, we reversed the tip sliding direction. Since the feedback was turned off during backward scanning, the gold on the tip apex was unavoidably worn out. Therefore, the SNOM images shown in Figs. 4(g)-4(i) of the main text were scanned by using different tips.

**Supplementary Figure 3.**

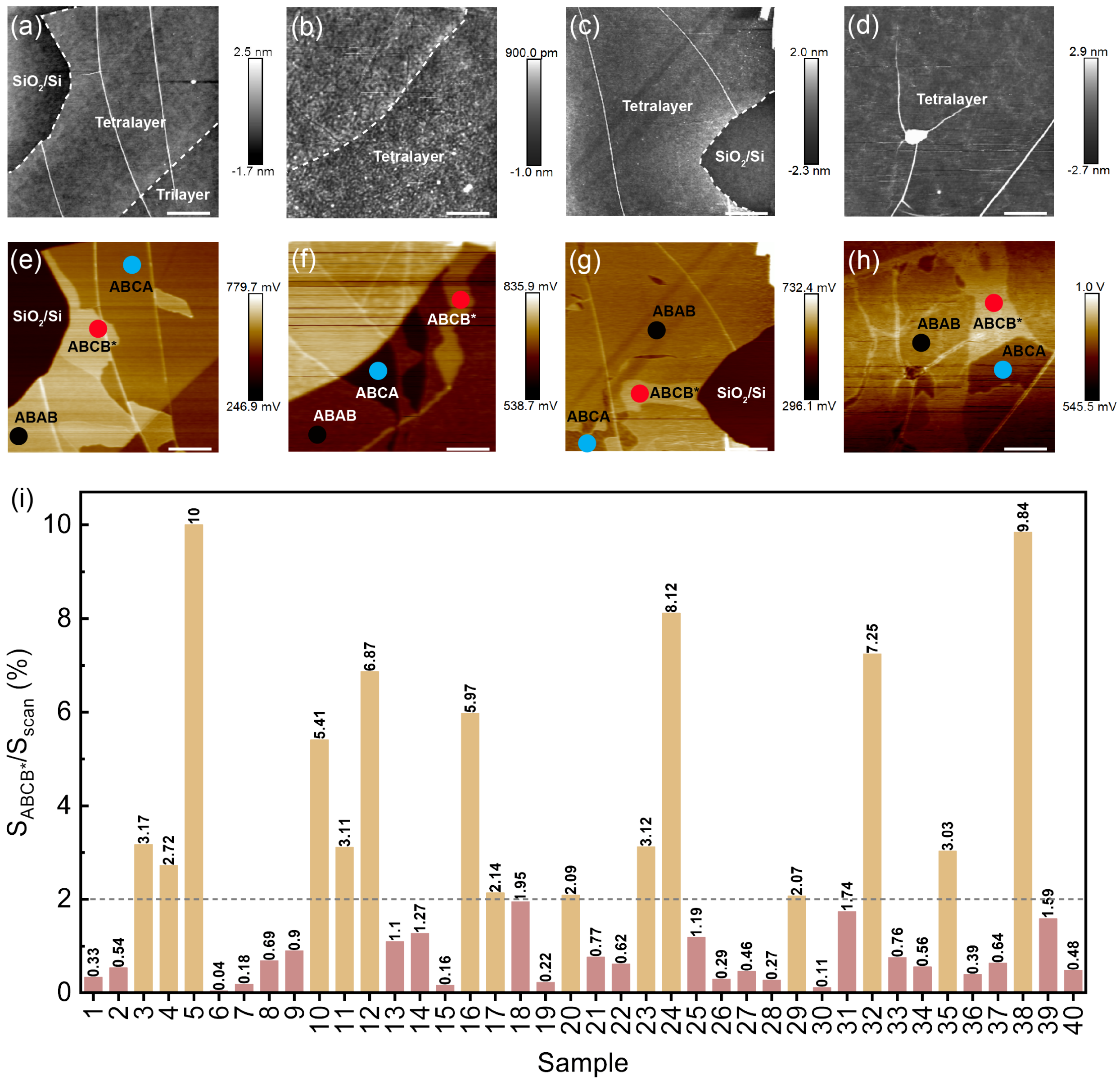

**FIG. S3. SNOM characterization of tetralayer graphene with nonpolar and polar domains.** (a)-(d) Topographies of four representative tetralayer graphene samples. The tetralayer regions are marked by white dashed lines. (e)-(h) SNOM images corresponding to the topographies in (a)-(d). The scale bars in (a) and (e) are 2 μm. The scale bars in (b), (d), (f), and (h) are 1 μm. The scale bars in (c) and (g) are 3 μm. (i) Histogram of the area portion of the ABCB-type (ABCB*) domains to the total scanning area for 40 tetralayer samples showing polar domains in this study. 15 samples have area proportions above 2% (yellow bars), whereas 25 samples have area proportions below 2% (red bars). The total number of scanned tetralayer samples is 132.

**Supplementary Note 3. Details of gate-dependent SNOM measurements of device 1**

As $V_g$ is increased from 0 V to 20 V [from Fig. 2(a) to Fig. 2(b) of the main text], a DW across the whole region becomes clearly visible, dividing the original region with nearly identical optical contrast into two parts, region I and region II, with slightly different optical contrasts, as shown in Fig. 2(b) of the main text. This DW becomes more apparent as $V_g$ increases to 40 V [Fig. 2(c) of the main text], where the DW separating region I and region II is indicated by the black arrow [16,17]. Figures 2(c) ($V_g$ = 40 V) and 2(d) ($V_g$ = 60 V) of the main text clearly show that as the gate voltage is continuously increased from $V_g$ = 20 V, the near-field response of region II becomes progressively darker than that of region I. The near-field response almost returns to its initial state [Fig. 2(a) of the main text] when $V_g$ passes through 0 V from the positive maximum, making regions I and II barely distinguishable [Fig. 2(e) of the main text]. This result implies that device 1 at $V_g$ = 0 V is close to its charge neutrality point [16,17]. When $V_g$ is gradually changed from $V_g$ = 0 V [Fig. 2(e) of the main text] to more negative values [Figs. 2(f)-2(h) of the main text], the near-field response of region I becomes progressively darker than that of region II. These figures thus show the exact opposite trend to that shown in Figs. 2(a)-2(d) of the main text with positive $V_g$. The complete dataset for the gate-dependent SNOM images is shown in Fig. S4. Consistent behaviors with those in Figs. 2(a)-2(h) of the main text (first cycle scan) are observed in a consecutive gate-dependent SNOM scan (second cycle scan) with extension to a larger range of $V_g$, from -60 V to 80 V (SNOM images shown in Fig. S5). Figure 2(i) of the main text shows that the two consecutive scan cycles reveal almost identical trends: (i) both the $(S_I - S_{II})/S_{II}$ and $(S_I - S_{III})/S_{III}$ optical contrasts reverse signs when the carrier type changes from electrons (positive $V_g$) to holes (negative $V_g$); (ii) the absolute optical contrast for both curves shows an increasing trend with increasing carrier doping for both electrons and holes.

**Supplementary Figure 4.**

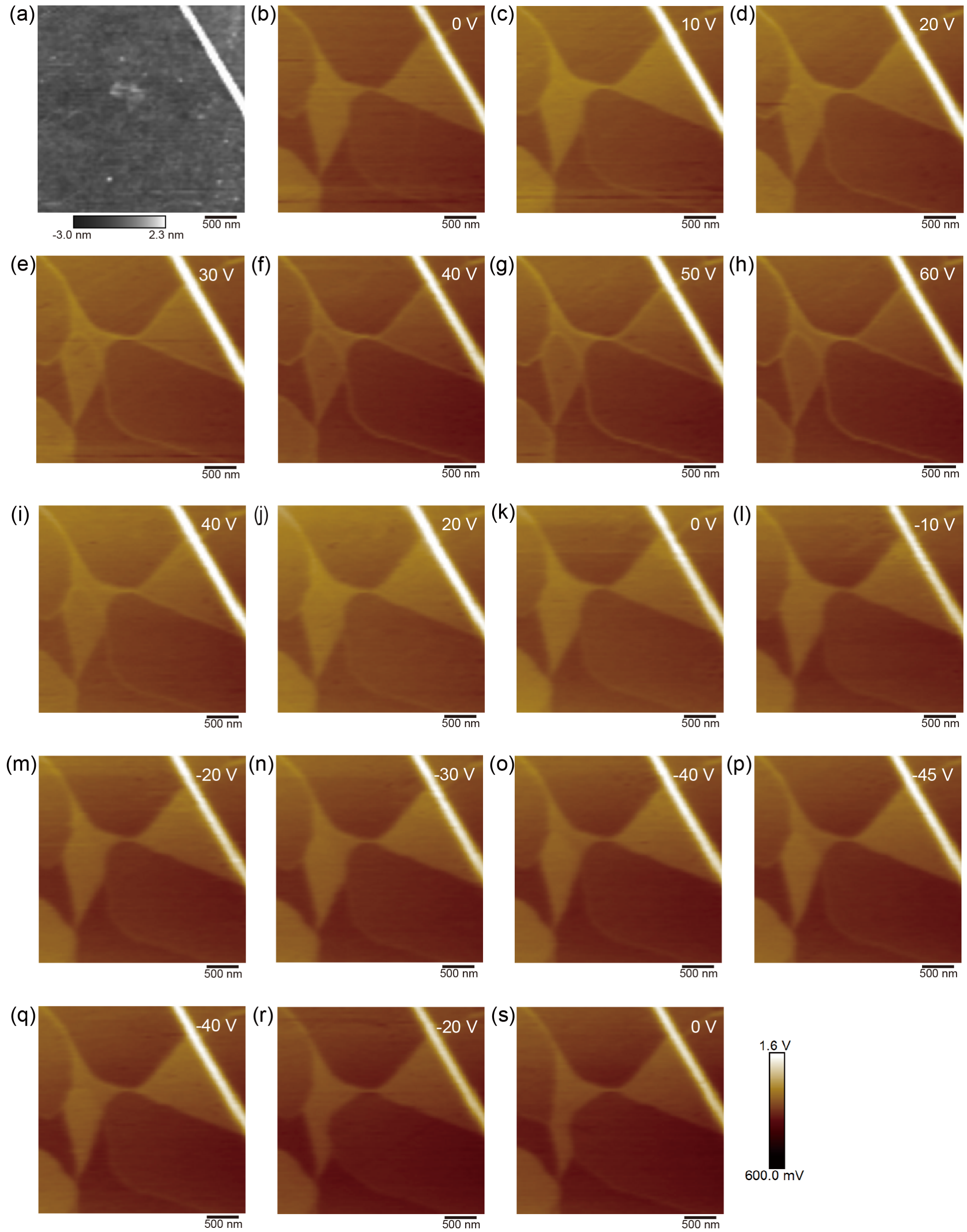

**FIG. S4. Full SNOM dataset for the first cycle scan of device 1.** (a) Topography of device 1. (b)-(s) SNOM images at different $V_g$ values following the applied $V_g$ sequence in our experiment. The applied $V_g$ ranges between -45 V and 60 V. All the presented figures share the same color bar (600 mV-1.6 V). (b), (d), (f), (h), (k), (m), (o), and (p) are presented in Figs. 2(a)-2(h) of the main text. The weak fringe features in the vertical direction of the SNOM images are presumably caused by the interference effect arising from light scattered from the nearby silver paste used to make sample contact. Figures 3(a)-3(c) in the main text is from this first cycle measurement (Fig. S4), and the corresponding figures are (n), (o), and (r).

**Supplementary Figure 5.**

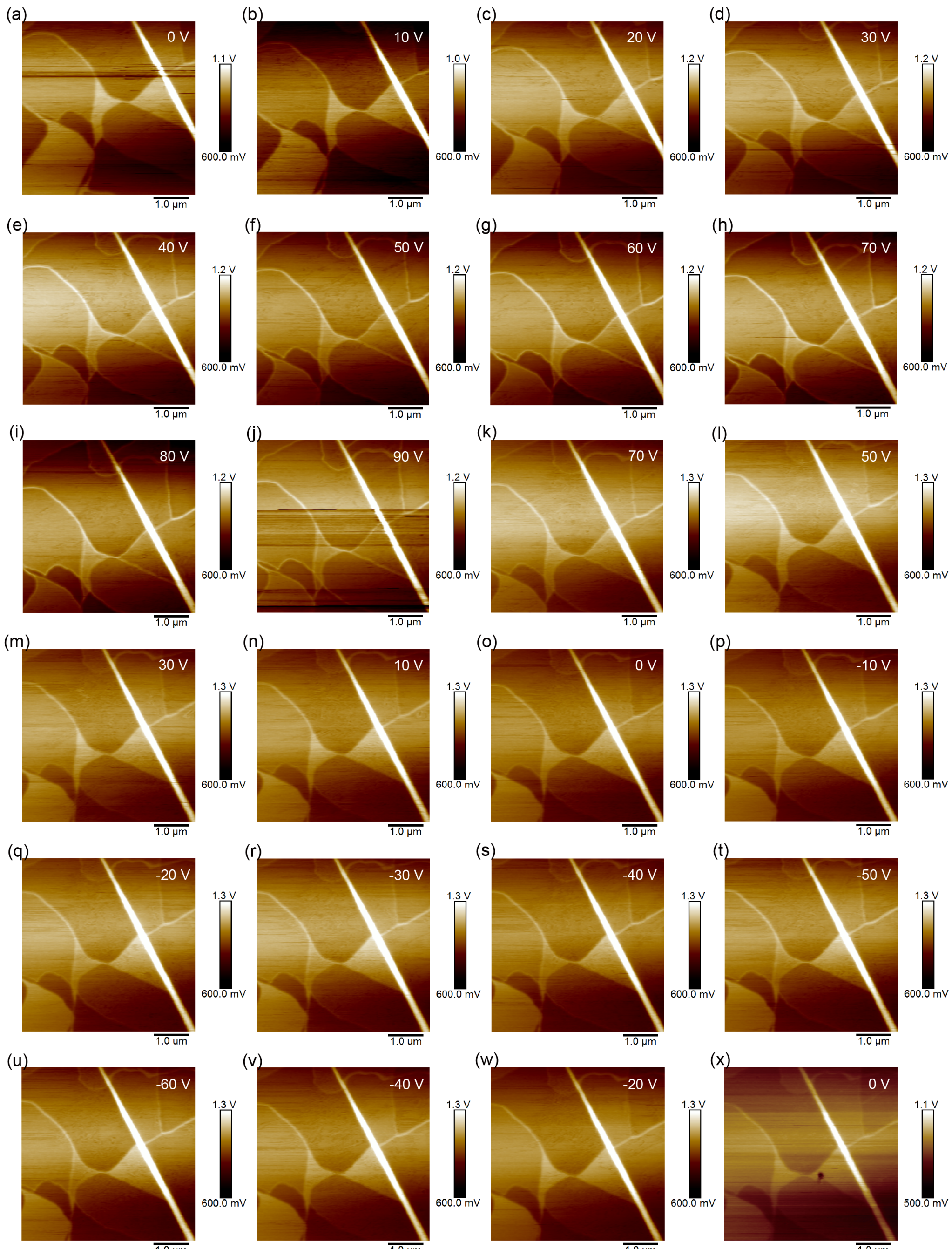

**FIG. S5. Full SNOM dataset for the second cycle scan of device 1.** (a)-(x) SNOM images at different $V_g$ values following the applied $V_g$ sequence in our experiment. The $V_g$ applied ranges between -60 V and 80 V, which is a larger range than that in the first cycle measurement shown in Fig. S4. The weak fringe features in the vertical direction of the SNOM images are presumably caused by the interference effect arising from light scattered from the nearby silver paste used to make sample contact. Figures 3(d) and 3(e) of the main text are from the second cycle measurement, and they correspond to (l) and (m) of Fig. S5, respectively. Both images in the main text share the same color scale bar from 600.0 mV to 1.3 V.

**Supplementary Figure 6.**

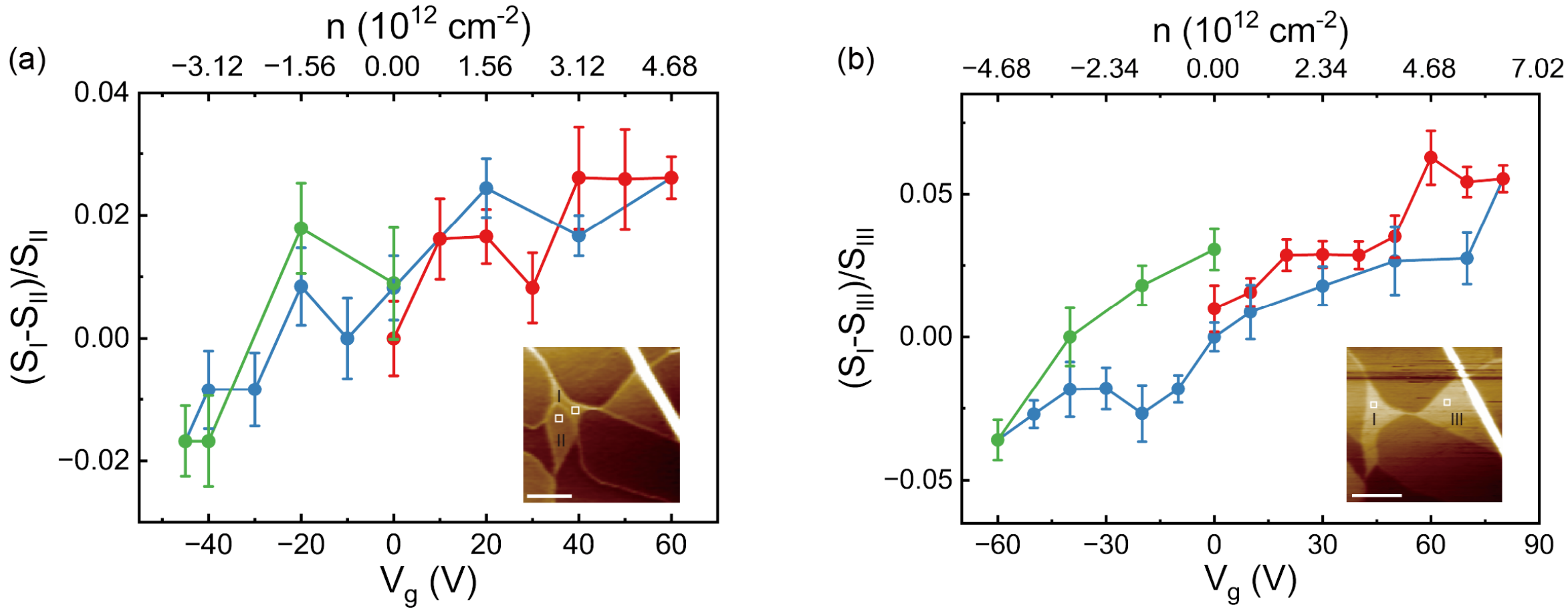

**FIG. S6. Full dataset for the optical contrast extracted from the two scan cycles of device 1.** (a) and (b) Full dataset for the optical contrast between the two polar stackings extracted from the first cycle (a) and second cycle (b) of gate-dependent SNOM measurements. In both (a) and (b), the extracted optical contrasts from $V_g = 0$ V to the positive maximum $V_g$ are denoted by red dots, those measured from the positive maximum $V_g$ to the negative maximum $V_g$ are denoted by blue dots, and those measured from the negative maximum $V_g$ back to $V_g = 0$ V are denoted by green dots. The results shown in Fig. 2(i) of the main text correspond to the data denoted by the red dots and the data denoted by the blue dots on the negative $V_g$ side. Insets in (a) and (b): the white boxes label the sample areas of the two nearby regions from which we extracted the average optical near-field response and obtained the optical contrast. The scale bars in the two insets are both 1 μm.

**Supplementary Figure 7.**

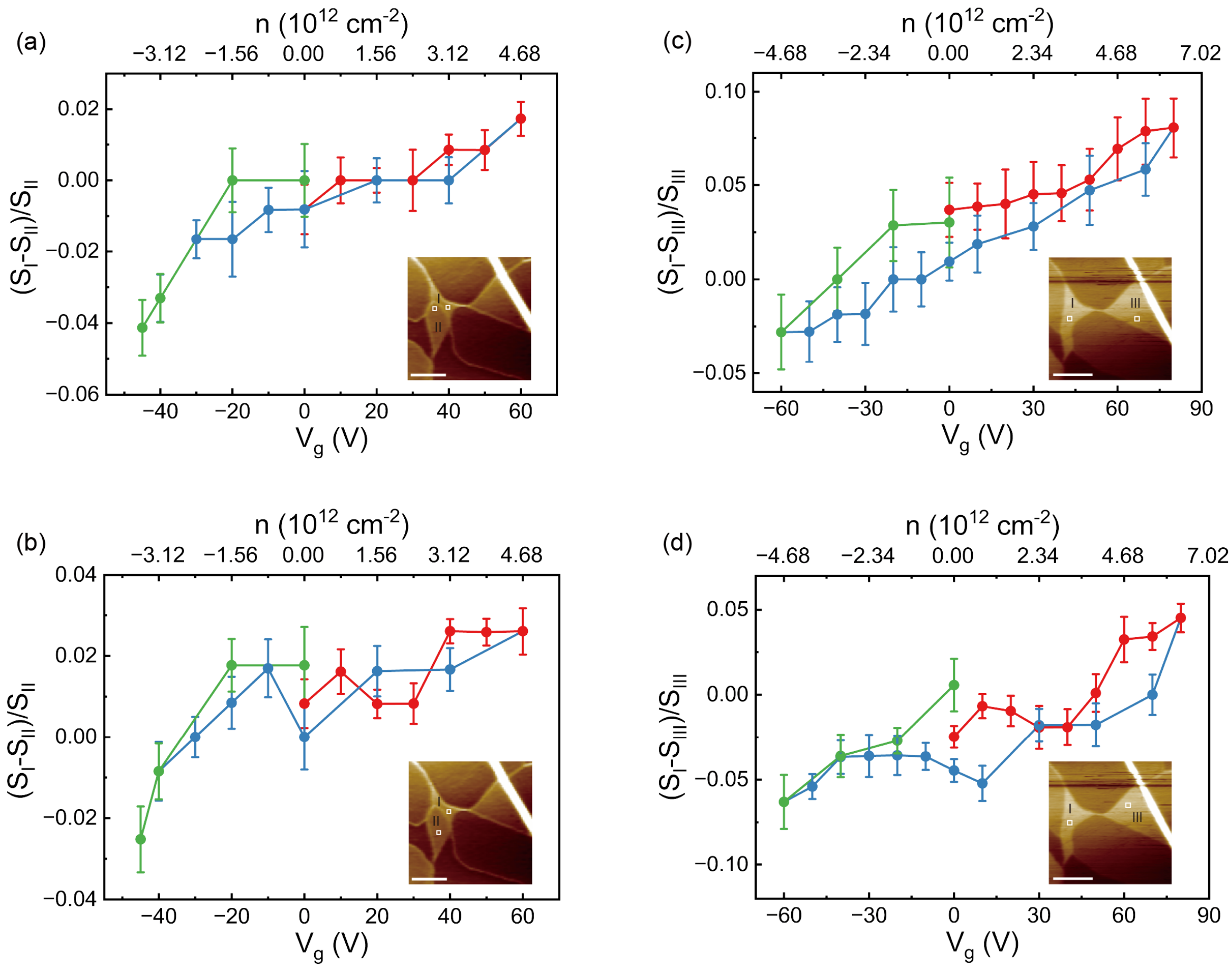

**FIG. S7. Examination of the extracted optical contrast between two polar stackings of device 1 by selecting different areas.** (a)-(d) Optical contrast between two polar stackings extracted along the same scanning rows [(a) from the first cycle and (c) from the second cycle] and different scanning rows [(b) from the first cycle and (d) from the second cycle] for the two cycles gate-dependent SNOM measurement. In (a) to (d), the extracted optical contrasts from $V_g$ = 0 V to positive maximum $V_g$ are denoted by red dots, those measured from positive maximum $V_g$ to negative maximum $V_g$ are denoted by blue dots, and those measured from negative maximum $V_g$ back to $V_g$ = 0 V are denoted by green dots. When comparing (a) with (b) and (c) with (d), we can observe that despite the selected areas are different, the trend of the extracted optical contrast upon varying $V_g$ is the same, consistent with those shown in Fig. 2(i) of the main text. Insets in (a) to (d): the white boxes label the sample areas from which we extract the average optical near-field response and obtain the optical contrast. The scale bars in all the insets are 1 μm.

**Supplementary Figure 8.**

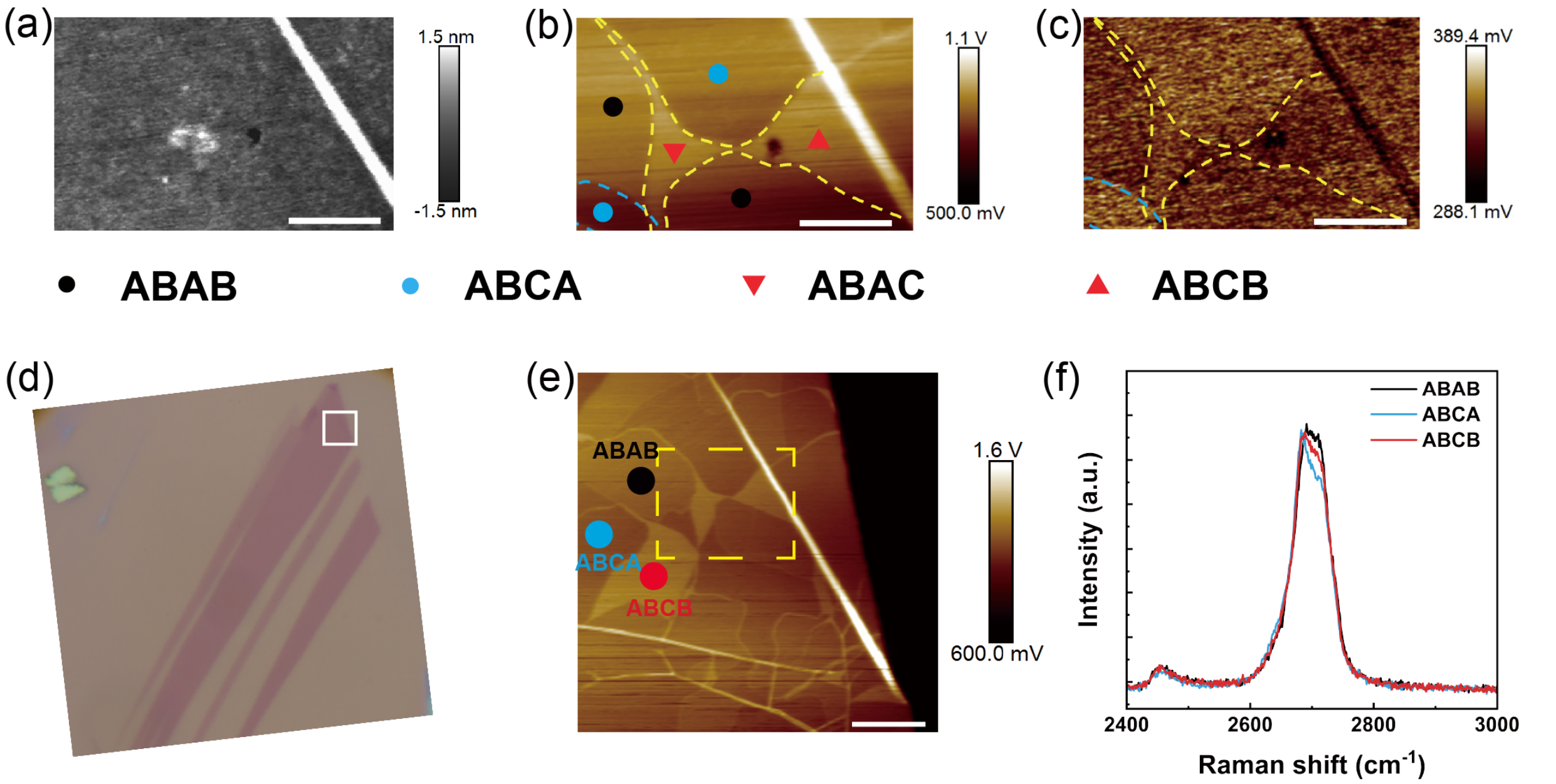

**FIG. S8. KPFM measurements and Raman spectra of device 1.** (a) Topography of the polar stackings in device 1. (b) SNOM image of the same tetralayer graphene area as in (a). Domains with distinct stackings are labeled by symbols of different colors. The dashed lines depict the contours of the DWs separating different stacking domains. (c) KPFM map of the same area as in (a) and (b) at $V_g = 0$ V. The measurement is carried out after two cycles of SNOM measurements (Figs. S4 and S5). The scale bars in (a) to (c) are all 1 µm. (d) Optical image of the tetralayer graphene. (e) SNOM image of the same area shown inside the white box in (d). The scanning area in Figs. 2(a)-2(h) of the main text coincides with the area enclosed by the yellow dashed frame. (f) Raman spectra obtained from the three different stackings, whose measurement points are marked in (e). Note that the small area of polar domains [at the center of the yellow dashed frame in (e)] makes direct far-field Raman measurements challenging. Instead, we measure the Raman spectra of the nearby polar domain with identical optical near-field responses [red dots in (e) with an identified ABCB stacking order]. Raman spectra collected from three distinct stackings exhibit noticeable differences, being consistent with the results shown in the main text [Fig. 1(e)] and with previous reports [10,11,22,23]. The scale bar in (e) is 2 µm.

**Supplementary Note 4. Correlating SNOM with KPFM.**

ABCB-type stacking orders are predicted to host vertical electric polarization because of the broken inversion and mirror symmetries [11,18-21]. The systematic evolution of the gate-dependent near-field response observed in Figs. 2(a)-2(i) of the main text suggests the existence of two distinct polar stackings hosting opposite vertical electric polarizations. Although opposite polar states were not observed in previous optical studies of tetralayer graphene [22,23], a recent KPFM study indicates their presence (but in isolation) in tetralayer graphene [12]. Our optical detection of two polar stacking orders with opposite vertical electric polarizations is further verified by our KPFM results [data shown in Fig. S8, alongside the Raman spectra supporting the stacking assignments marked in Fig. 2(b) of the main text], obtained after the two consecutive SNOM measurements [Fig. 2(i) of the main text] without applying a back-gate voltage. Distinct work functions are observed between two adjacent polar stacking orders, regions I and III (region II vanishes after the SNOM scans), in which the measured work function of region I is smaller than that of region III. Therefore, region I corresponds to ABAC stacking with downward polarization on the basis of our DFT calculations [Fig. S1(b) and Fig. S2]; while regions II and III, with larger work functions, correspond to ABCB stacking with upward polarization [Fig. S1(d) and Fig. S2].

**Supplementary Figure 9.**

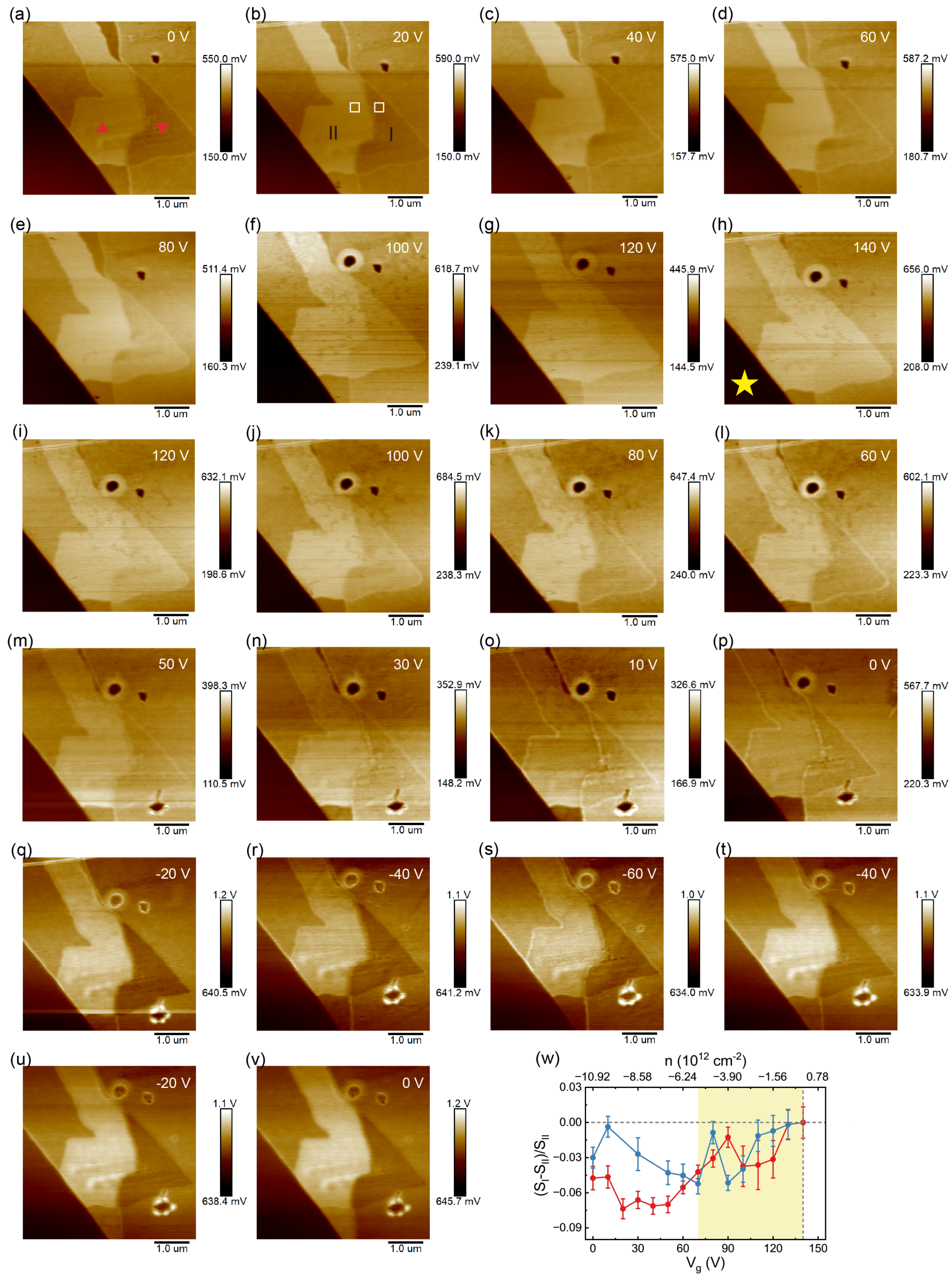

**FIG. S9. Full dataset for gate-dependent SNOM of device 2.** (a)-(v) SNOM images at different $V_g$ values following the applied $V_g$ sequence in our experiment between -60 V and 140 V. The optical near-field response between the two adjacent polar states becomes almost identical at approximately $V_g$ = 140 V (h), indicating strong hole doping in the sample at $V_g$ = 0 V, which is presumably due to unintentional doping from the substrate. Owing to this unintentional hole doping, we can identify ABAC and ABCB stackings without gating, and electron doping of the sample is not achieved. (w) Optical contrast between two adjacent polar stackings as a function of $V_g$. Red and blue dots mark extracted optical contrasts measured from $V_g$ = 0 V to the positive maximum $V_g$ and the positive maximum $V_g$ to the negative maximum $V_g$, respectively. The charge neutrality point roughly lies at $V_g$ = 140 V (indicated by the gray dashed lines), confirmed by the vanishing amplitude of the plasmon fringes reflected by the DW. The yellow shaded area marks the region from a carrier density of 0 to a hole doping density of $5.46 \times 10^{12} cm^{-2}$, in which the optical contrast tends to increase with increasing hole doping. The white boxes in (b) label the sample areas from which the optical contrasts were extracted.

**Supplementary Figure 10.**

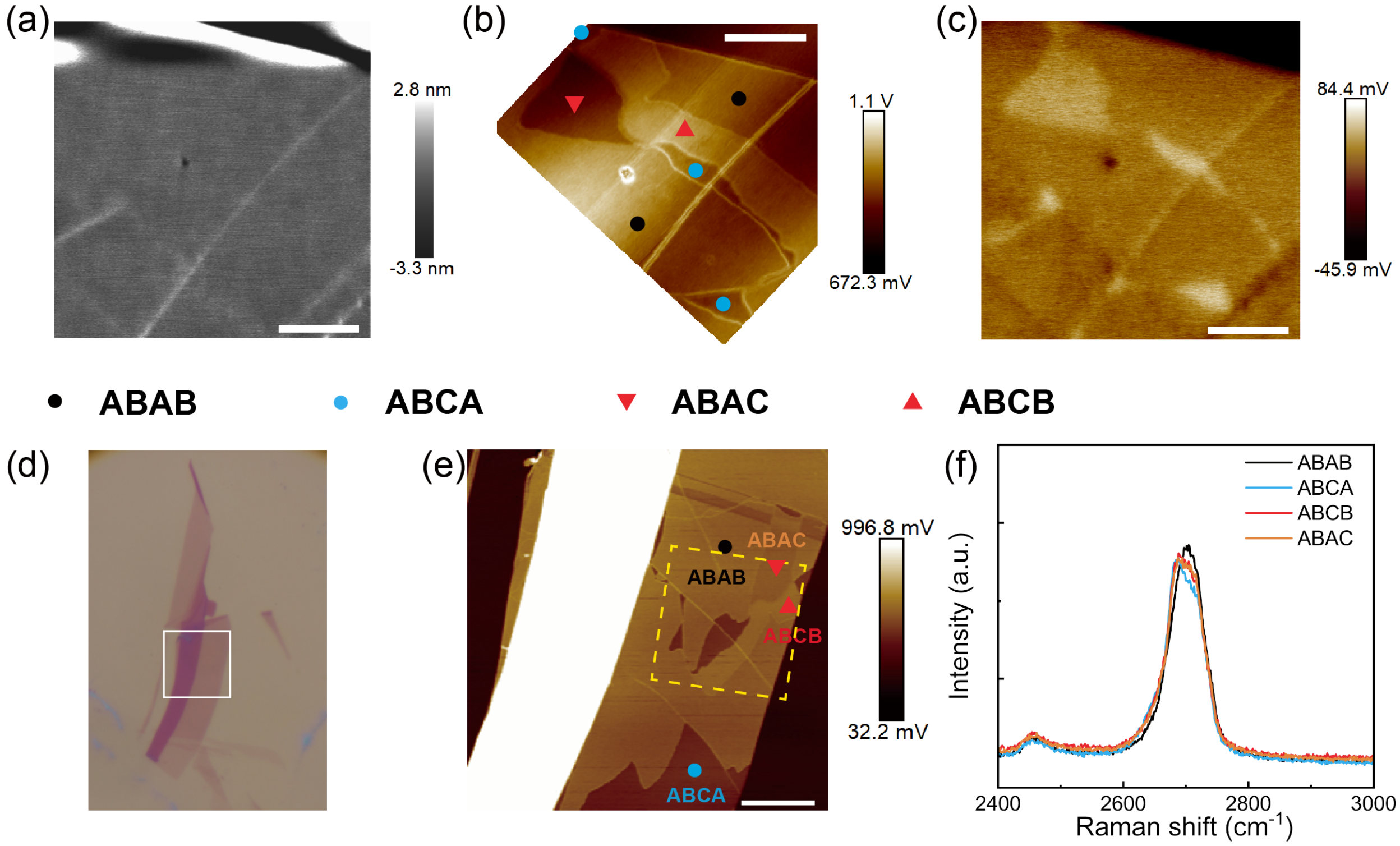

**FIG. S10. KPFM measurements and Raman spectra of device 2.** (a) Topography of device 2. (b) SNOM image of the same area as in (a). Domains with distinct stackings are labeled by symbols of different colors. At $V_g$ = 0 V, the two polar stackings (indicated by triangles) have different near-field optical contrasts since this sample is hole-doped at $V_g$ = 0 V. (c) KPFM mapping of the same area as in (a) and (b) at $V_g$ = 0 V. The KPFM results of device 2 are consistent with those of device 1 (Fig. S8), device 3 (Fig. S12), and device 4 (Fig. S15). The scale bars in (a) to (c) are all 2 μm. (d) Optical image of the tetralayer graphene. The white box marks the scanning area of (e). (e) SNOM image of the tetralayer graphene area shown inside the white box in (d), which contains four different optical near-field contrasts at $V_g$ = 0 V. The yellow dashed frame labels the scanning area of KPFM mapping. (f) Raman spectra of the four different stackings. The different symbols in (e) mark the positions where the Raman spectra in (f) were taken. The scale bar in (e) is 4 μm.

**Supplementary Figure 11.**

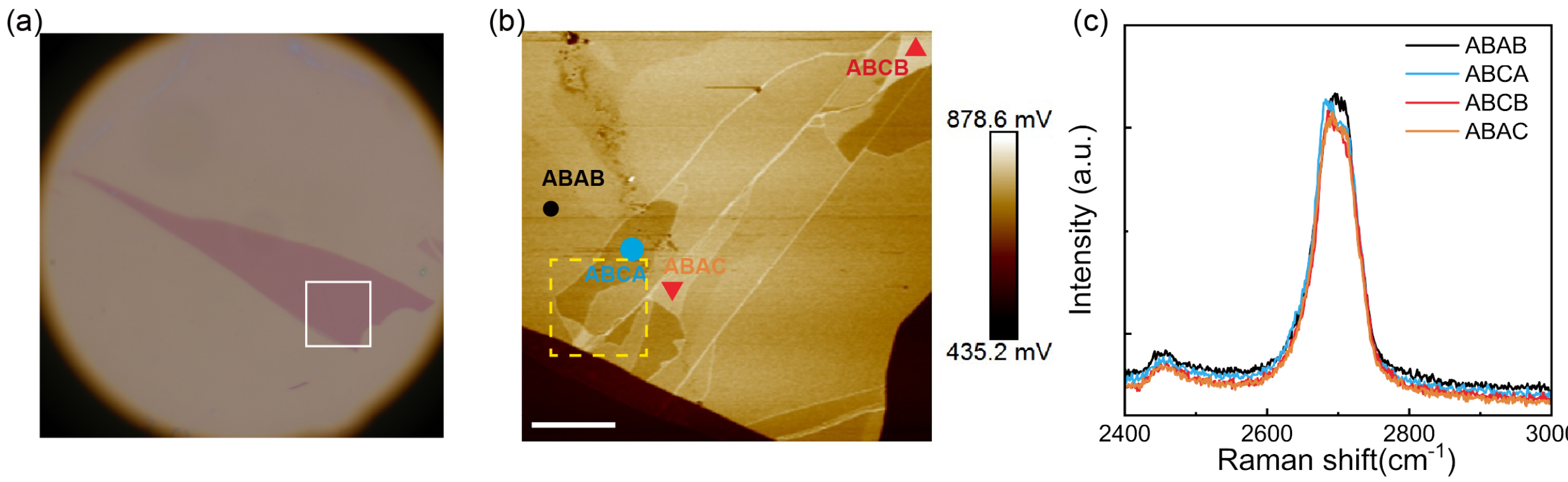

**FIG. S11. Raman spectra of device 3.** (a) Optical image of the tetralayer graphene. The white box marks the scanning area of (b). (b) SNOM image of the same tetralayer graphene area shown inside the white box in (a), which contains four different optical near-field contrasts at $V_g = 0$ V. The yellow dashed frame labels the scanning area of KPFM mapping shown in Fig. S12. (c) Raman spectra of the four different stackings. Different symbols in (b) mark the positions where the Raman spectra in (c) are taken. Note that the small area of polar domains [at the corner of the yellow dashed frame in (b)] makes direct far-field Raman measurements challenging. We instead measure the Raman spectra of the nearby polar domain that shows an identical optical near-field response [red triangles in (b)]. The scale bar in (b) is 4 μm.

**Supplementary Figure 12.**

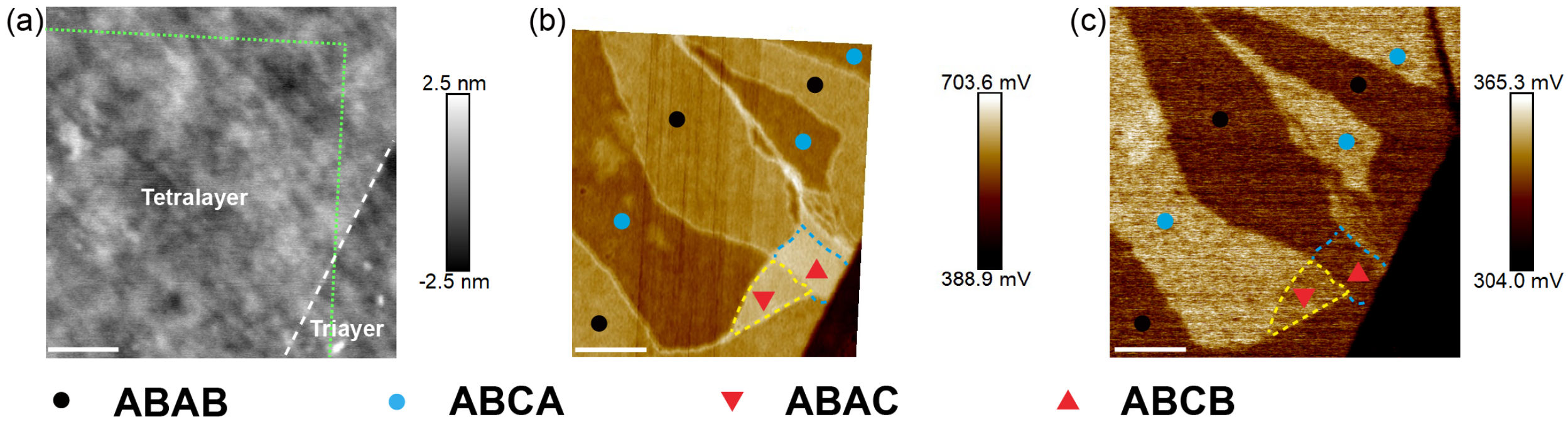

**FIG. S12. KPFM measurement of device 3 at $V_g$ = 0 V.** (a) Topography of the tetralayer graphene. (b) SNOM image of the green dashed area in (a). Domains with distinct stackings are labeled by different symbols in color. At $V_g$ = 0 V, the two polar stackings (indicated by triangles) show different near-field optical contrast, since this sample is hole-doped at $V_g$ = 0 V. The stacking assignments in (b) are also supported by our Raman spectra in Fig. S11 and the KPFM result in (c). (c) KPFM mapping of the same area as in (a) at $V_g$ = 0 V. For the two nearby polar stackings, the one that is darker (brighter) in SNOM image appears to be brighter (darker) in KPFM, indicative of the downward polarization [downward triangle in (b) and (c)] and upward polarizations [upward triangle in (b) and (c)], respectively. The blue dashed line marks the boundary of the ABCB domain, and the yellow dashed line marks the boundary of the ABAC domain. The KPFM results of device 3 are consistent with those of device 1 (Fig. S8), device 2 (Fig. S10), and device 4 (Fig. S15). All the scale bars are 1 μm.

**Supplementary Figure 13.**

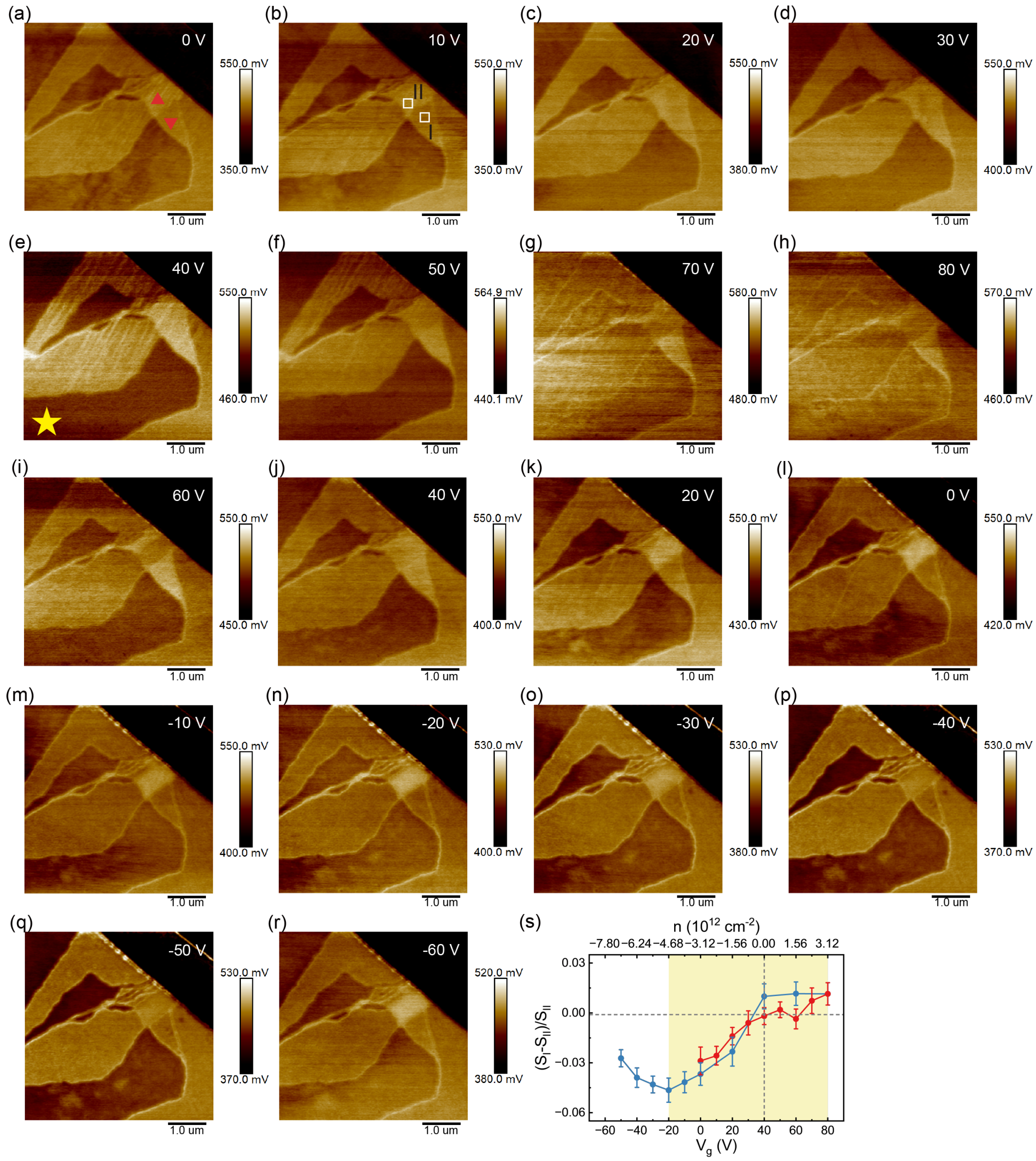

**FIG. S13. Full dataset for the gate-dependent SNOM of device 3.** (a)-(r) SNOM images at different $V_g$ following the sequence of applied $V_g$ in our experiment between -60 V and 80 V. The optical near-field response between the two adjacent polar states becomes almost identical at around $V_g = 40$ V (e), indicating hole doping in the sample at $V_g = 0$ V. This means that the sample is slightly hole-doped at $V_g = 0$ V, which prevents our tetralayer from being further electron-doped. (s) Optical contrast between the two adjacent polar stackings, as a function of $V_g$. Red/blue dots mark extracted optical contrasts measured from $V_g$ = 0 V to the positive maximum $V_g$ and the positive maximum $V_g$ to the negative maximum $V_g$, respectively. The charge neutrality point roughly lies at $V_g$ = 40 V (indicated by the gray dashed lines), confirmed by the vanishing amplitude of the plasmon fringes reflected by the DW. The yellow shaded area marks the carrier density from electron doping density of $3.12 \times 10^{12} cm^{-2}$ to hole doping density of $\mathbf{4}.68 \times 10^{12} cm^{-2}$, where the optical contrast shows an increased trend with increasing doping. The white boxes in (b) label the sample areas in which the optical contrasts are extracted.

**Supplementary Figure 14.**

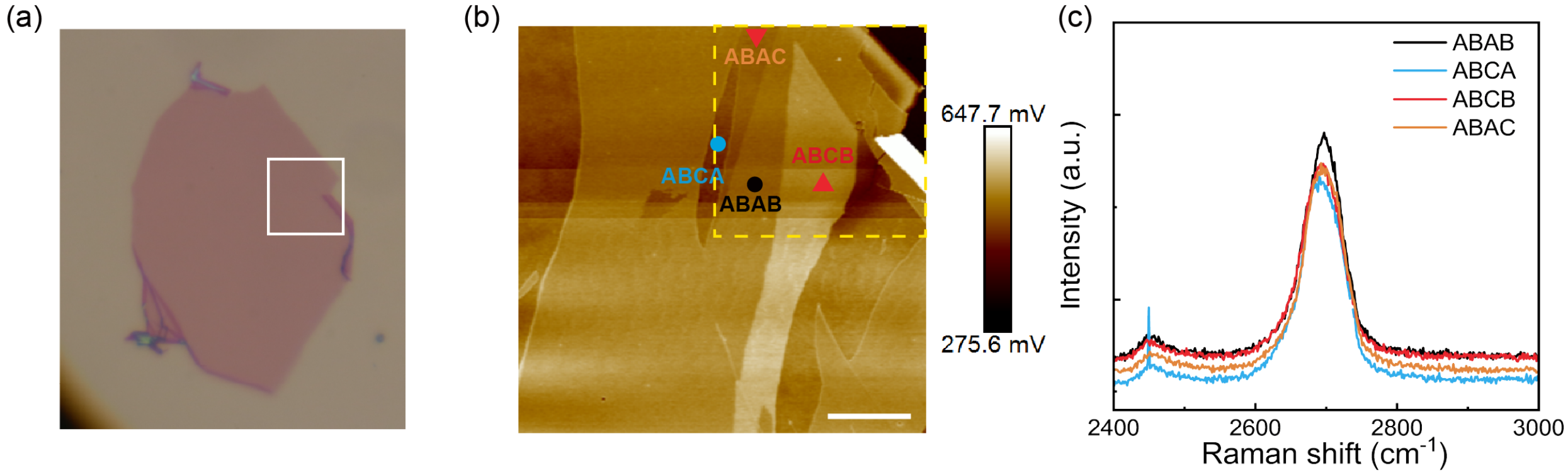

**FIG. S14. Raman spectra of device 4.** (a) Optical image of the tetralayer graphene. The white box marks the scanning area of (b). (b) SNOM image of the same tetralayer graphene area shown inside the white box in (a), which contains four different optical near-field contrasts at $V_g = 0$ V. The yellow dashed frame labels the scanning area of KPFM mapping shown in Fig. S15. (c) Raman spectra of the four different stackings. Different symbols in (b) mark the positions where the Raman spectra in (c) are taken. To better distinguish the spectra of different stackings, we manually shift the spectrum of ABCA stacking. The scale bar in (b) is 6 μm.

**Supplementary Figure 15.**

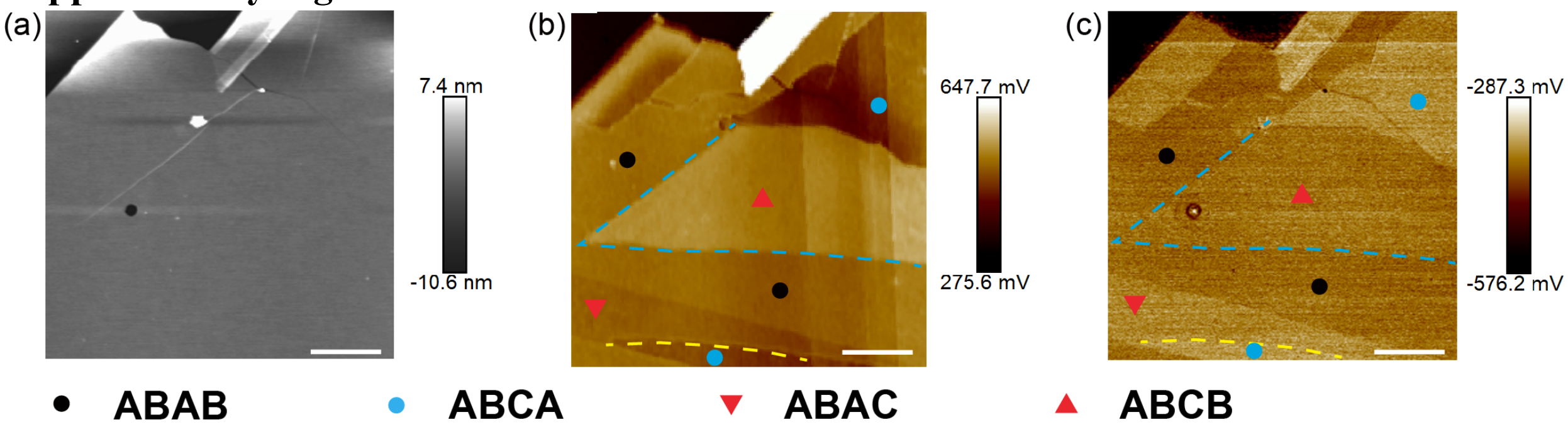

**FIG. S15. KPFM measurement of device 4 at $V_g$ = 0 V.** (a) Topography of the tetralayer graphene. (b) SNOM image of the same area as in (a). Domains with distinct stackings are labeled by different symbols in color. At $V_g = 0$ V, the two polar stackings (indicated by triangles) show different near-field optical contrast, since this sample is hole-doped at $V_g = 0$ V. The stacking assignments in (b) are also supported by our Raman spectra in Fig. S14 and the KPFM result in (c). (c) KPFM mapping of the same area at $V_g = 0$ V. For the two nearby (not directly being adjacent) polar stackings, the one that is darker (brighter) in SNOM image appears to be brighter (darker) in KPFM, indicative of the downward polarization [downward triangle in (b) and (c)] and upward polarizations [upward triangle in (b) and (c)], respectively. The blue (yellow) dashed line marks the boundary between the ABAB domain and the ABCB domain (between the ABCA domain and the ABAC domain), respectively. The KPFM results of device 3 are consistent with those of device 1 (Fig. S8), device 2 (Fig. S10), and device 3 (Fig. S12). All the scale bars in (a), (b), and (c) are 3 μm.

**Supplementary Figure 16.**

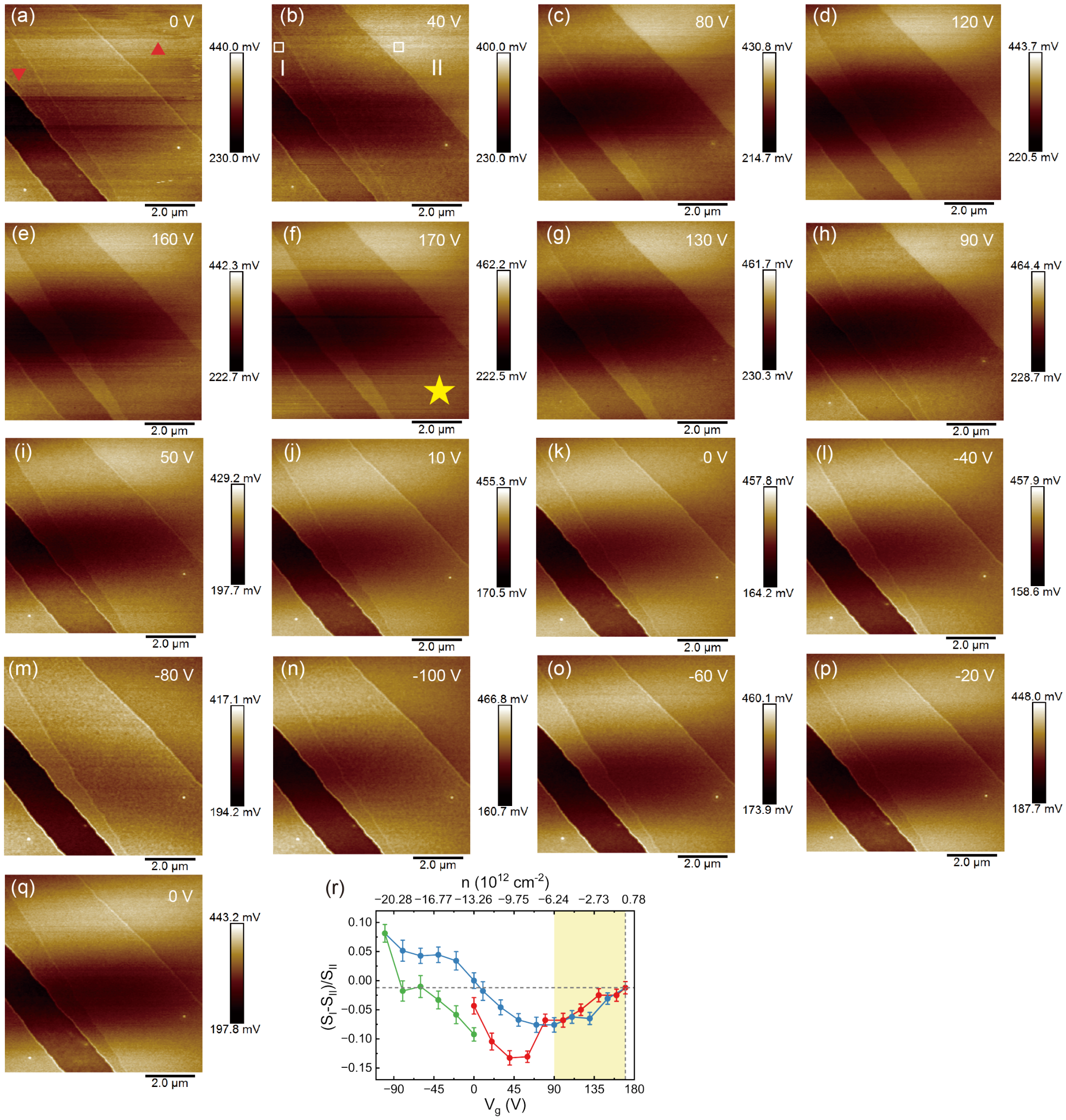

**FIG. S16. Full dataset for the gate-dependent SNOM of device 4.** (a)-(q) SNOM images at different $V_g$ following the sequence of applied $V_g$ in our experiment, between -100 V and 170 V. The optical near-field response between the two nearby polar states becomes almost identical at around $V_g = 170$ V (f), indicating a strong hole doping at $V_g = 0$ V, which prevents our tetralayer from being electron-doped. We note that fringe features appearing in the vertical direction of SNOM images are presumably due to the interference effect caused by the scattered light from the nearby silver paste used to make sample contact. (r) Optical contrast between the two nearby polar stackings, as a function of $V_g$. Red/blue/green dots mark optical contrasts along $V_g = 0$ V to positive maximum $V_g$, positive maximum $V_g$ to negative maximum $V_g$, and negative maximum $V_g$ to $V_g =$ 0 V trajectories, respectively. The charge neutrality point roughly lies at = 170 V (indicated by the gray dashed lines), confirmed by the vanishing amplitude of the plasmon fringes reflected by the DW. The yellow shaded area marks the carrier density from 0 to hole doping density of $6.24 \times 10^{12} cm^{-2}$, where the optical contrast shows an increasing trend with increasing hole doping. The white boxes in (b) label the sample areas in which the optical contrasts are extracted.

**Supplementary Figure 17.**

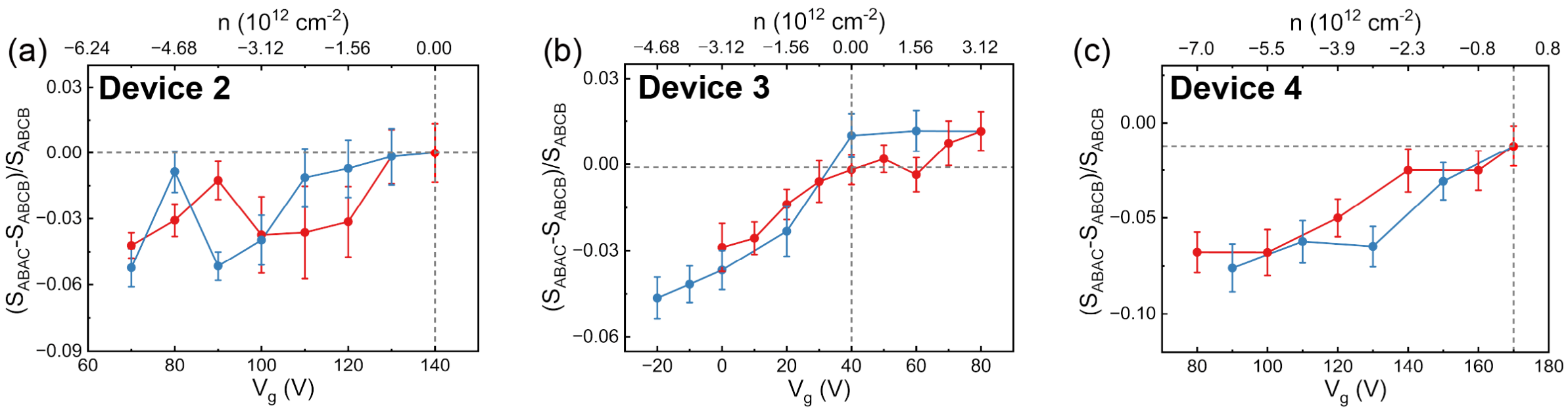

**FIG. S17. Extracted optical contrast between the two polar stackings of devices 2-4 (within a certain carrier density range).** (a)-(c) Optical contrasts between two adjacent polar stackings of devices 2-4 as a function of $V_g$ (carrier density $n$). The optical contrast in devices 2-4 tends to increase with increasing doping, corresponding to the yellow shaded areas in Fig. S9(w), Fig. S13(s), and Fig. S16(r). The dashed lines in (a) to (c) indicate the charge neutrality points in the three devices. The gate-dependent optical contrasts for the forward scan (increasing $V_g$) and backward scan (decreasing $V_g$) are denoted by the red and blue dots, respectively.

**Supplementary Note 5. Theoretical interpretation of the gate-dependent near-field response of polar stackings**

If we assume that the detected near-field signal underneath the tip is predominantly contributed by the free carriers residing in the topmost graphene layer (closest to the tip), the observed distinct response in Fig. 2(i) of the main text can be understood as follows. The equal numbers of purely stacking-induced electrons and holes in the topmost graphene layer in ABAC and ABCB obtained from DFT calculations according to the calculated charge density distribution shown in Figs. S1(b) and S1(d) and the calculated partial Löwdin charge in Fig. S2. It is schematically illustrated in the top panel in Fig. S18(a), which makes the near-field optical responses in ABAC and ABCB identical at $V_g = 0$ V [top panel in Fig. S18(a)] [24,25]. When $V_g > 0$ (electron doping), doped electrons in ABAC are mainly filled in the fourth layer (L4), and virtually no electrons are filled in the first layer (L1) according to the calculated layer-projected density of states (LDOS) [Fig. S18(b)]. Therefore, the carrier concentration in the topmost layer basically remains unchanged [middle panel of Fig. S18(a)] in ABAC. In contrast, most of the electrons in ABCB are filled in the first layer (L1) when $V_g > 0$, as indicated in Fig. S18(c). Since the dominant carriers in the topmost graphene layer are holes for the ABCB stacking due to the intrinsic charge distribution [top panel in Fig. S18(a)], electron filling decreases the carrier concentration in the topmost layer [middle panel of Fig. S18(a)]. As a result, the ABCB stacking appears to become increasingly darker than the ABAC stacking with increasing electron doping. The trend for $V_g < 0$ (hole doping) can be similarly understood [schematically illustrated in the bottom panel of Fig. S18(a), according to the LDOS in Figs. S18(b) and S18(c)]. We note that our DFT calculations in Figs. S18(b) and S18(c) do not consider the gating field-induced change, which may modify the charge distribution among different layers in tetralayer graphene.

**Supplementary Figure 18.**

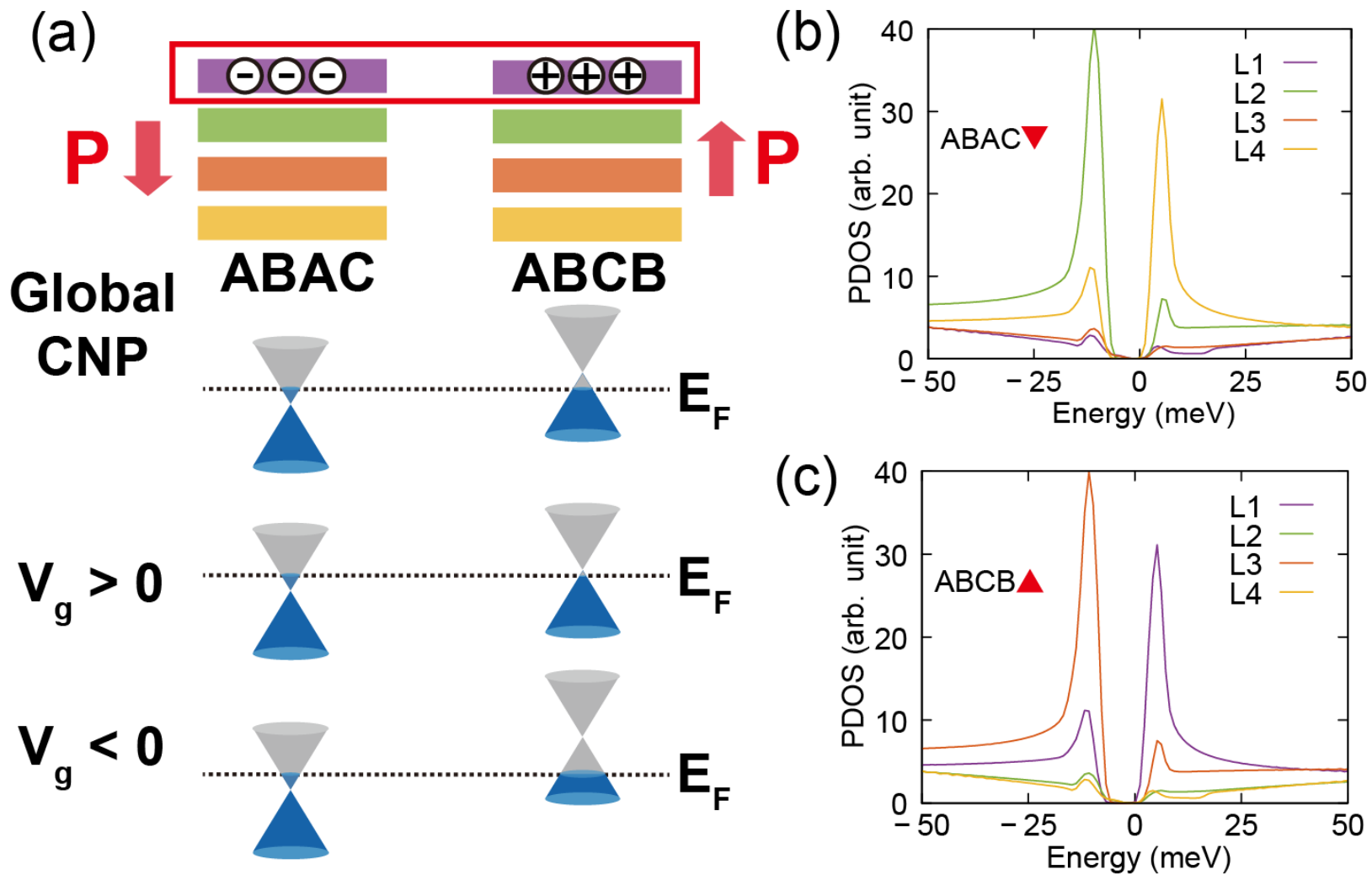

**FIG. S18. Theoretical interpretation of the gate-dependent optical contrast of polar stacking.** (a) Schematic of the proposed model explaining the observed optical contrast change in Fig. 2(i) of the main text. The carriers of the topmost graphene layer are electrons and holes for ABAC (polarization downward) and ABCB (polarization upward) stackings, respectively, at the global charge neutrality point (CNP) according to the calculated charge density distribution shown in Figs. S1(b) and S1(d). The filling of the topmost graphene layer under gating follows the calculated layer-projected DOSs in (b) and (c). (b),(c) Calculated layer-projected DOSs for ABAC and ABCB stackings at CNP, respectively.

**Supplementary Figure 19.**

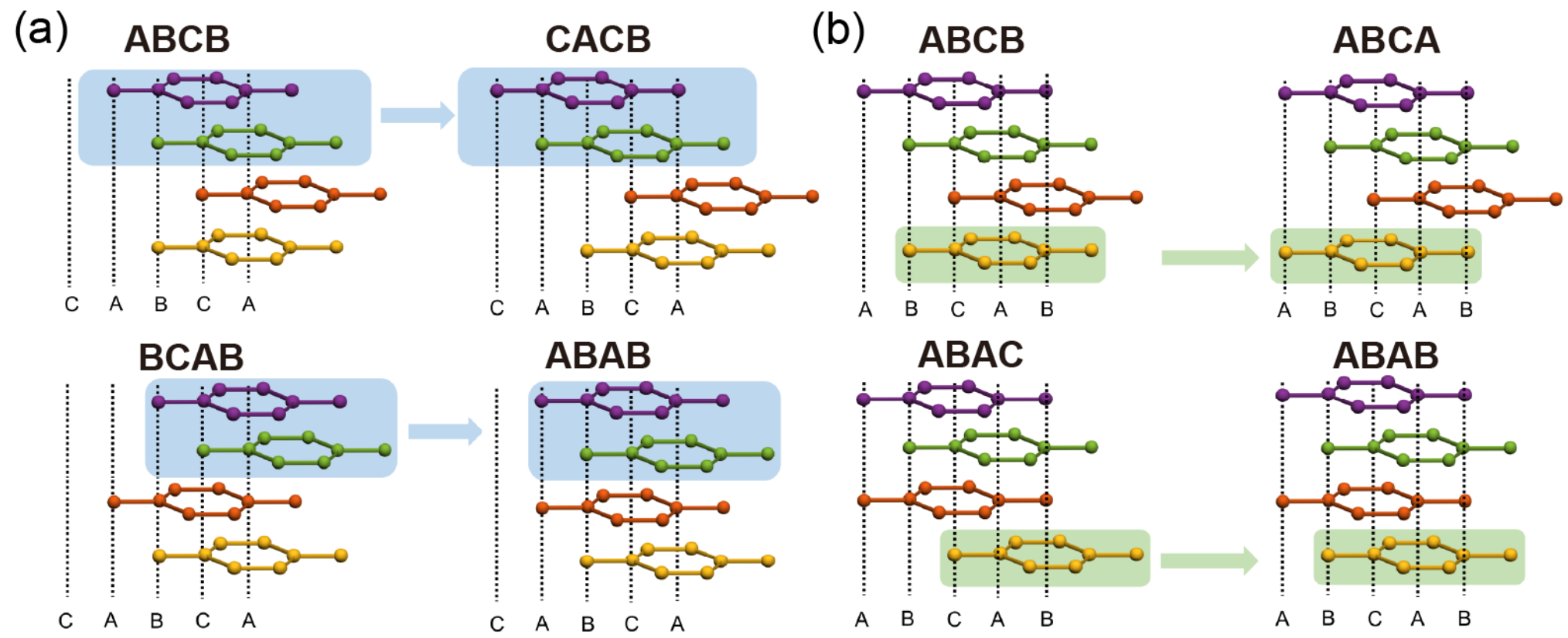

**FIG. S19. Schematic illustration of a second possibility regarding the interlayer sliding.** (a) Schematics of the second possibility, in which sliding the top two layers together also corresponds to the blue DW motion in Fig. 3(f) of the main text. Note that ABCB and CABA are equivalent stacking orders, and BCAB and ABCA are equivalent stacking orders. (b) Schematics of the second possibility, in which sliding the bottom layer only also corresponds to the green DW motion in Fig. 3(f) of the main text. Note that ABCB and CABA are equivalent stacking orders, and ABAC and CACB are equivalent stacking orders.

**Supplementary Figure 20.**

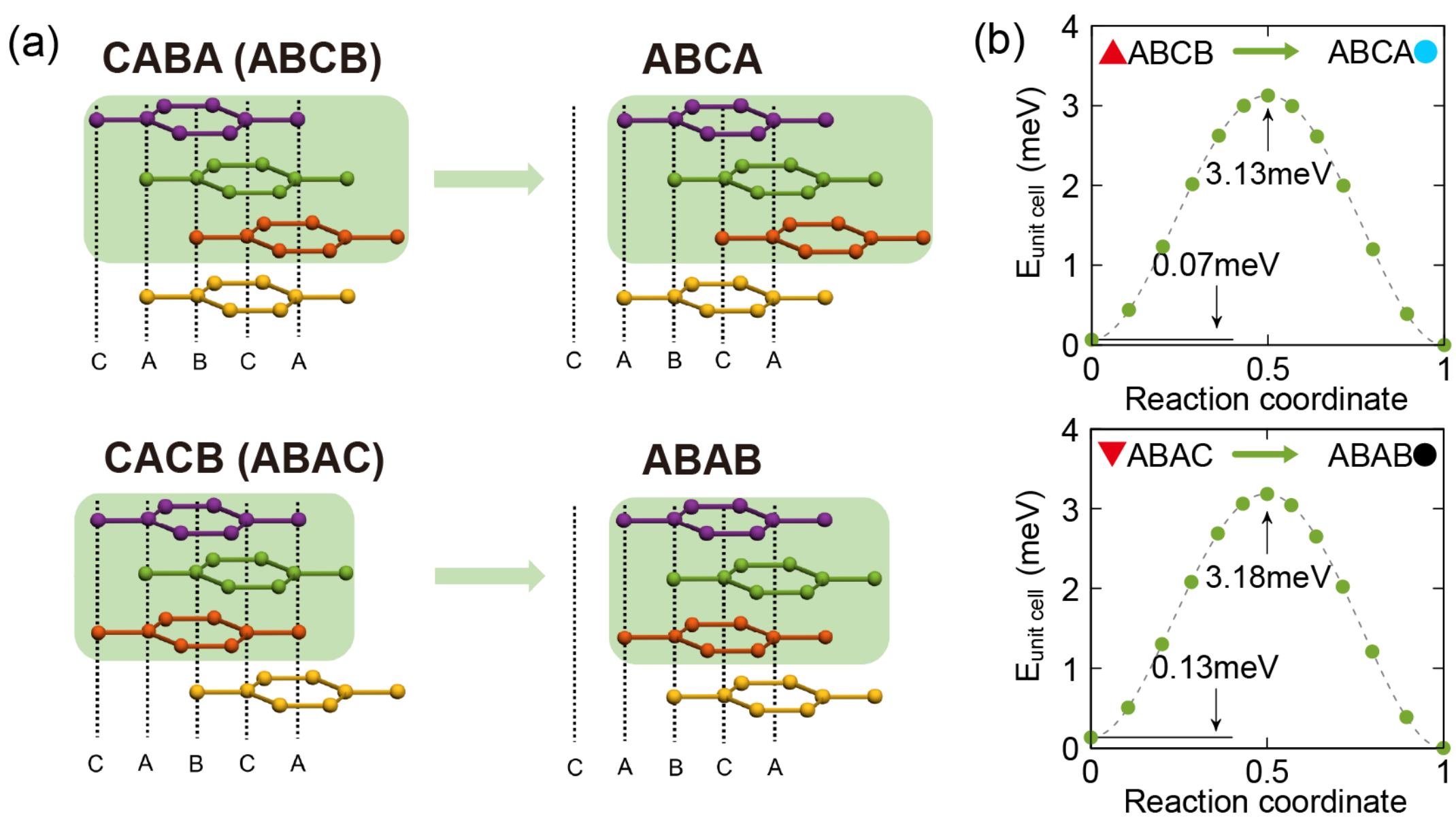

**FIG. S20. Schematic illustration and the calculated energy barriers of the green domain wall sliding in Fig. 3(f) of the main text.** (a) Schematic illustration for the green DW motion in Fig. 3(f) of the main text, corresponding to sliding of the top three layers together [changing from CABA(ABCB) to ABCA and from CACB (ABAC) to ABAB]. (b) Calculated total energies for ABAB, ABCA, ABAC, and ABCB stackings and energy barriers for the two switching pathways corresponding to the green DW sliding observed in Figs. 2(a)-2(e) of the main text.

**Supplementary Figure 21.**

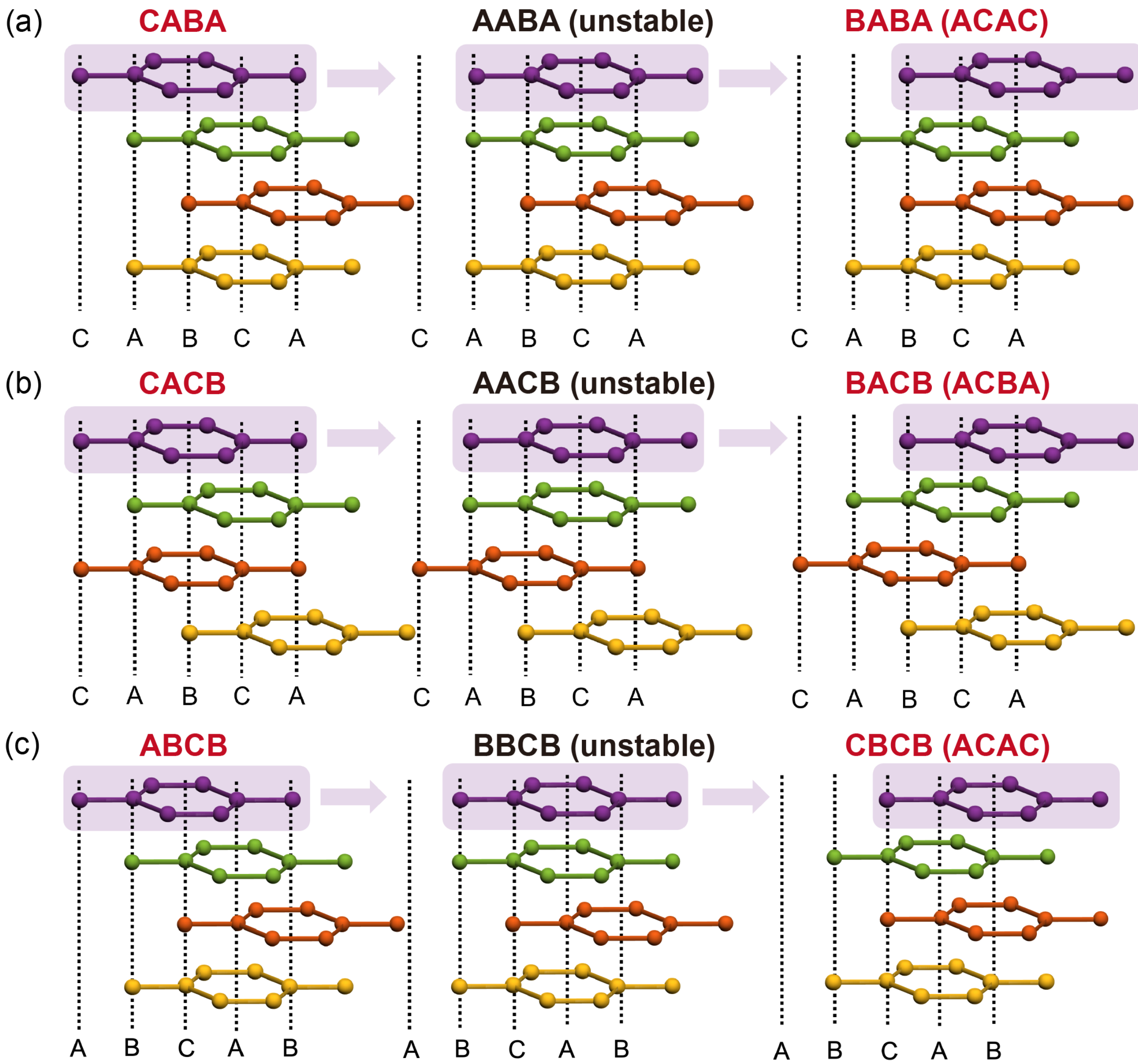

**FIG. S21. Top-layer sliding cannot account for green DW motion in Fig. 3(f) of the main text.** (a)-(c) Possible stackings obtained by sliding the topmost layer starting from CABA, CACB, and ABCB stackings, respectively. The domains with AA and BB adjacent stackings are unstable. The movement of the green DW cannot be achieved by sliding the topmost layer.

**Supplementary Note 6. Calculated total energies for different stackings and energy barriers for four switching pathways**

According to the DFT-calculated switching pathways presented in Fig. 3(h) of the main text and Fig. S20(b), the nonpolar ABCA and two polar stackings (which are energetically degenerate, as they should be) are clearly metastable phases, with total energies 0.07 meV/unit cell and 0.13 meV/unit cell higher than that of ABAB, respectively. The calculated total energy difference in Fig. 3(h) of the main text and Fig. S20(b) is qualitatively consistent with previous reports [18,22,23]. According to Fig. 3(h) of the main text and Fig. S20(b), although different stackings have slightly different thermodynamic stabilities, the calculated energy barriers for the four interlayer sliding pathways that characterize the two types of DW motion observed in the experiment [blue and green DWs and domain patterns in Fig. 3(f) of the main text] vary only slightly [Fig. 3(h) of the main text and Fig. S20(b)], implying that there is little preference among the four pathways in terms of the energy barrier to surmount. This result agrees with our observations in Figs. 3(a)-3(f) of the main text that the two types of DWs both occur under gating without a noticeable preference.